\documentclass[journal,onecolumn,draft]{IEEEtran}
\pdfoutput=1

\usepackage{cite}

\ifCLASSINFOpdf
   \usepackage[pdftex]{graphicx}
\else
\fi
\usepackage{amsmath}
\usepackage{amssymb}  
\usepackage{stmaryrd}
\usepackage{color}

\newcommand{\len}{L_c}

\begin{document}

\title{The Velocity of the Propagating Wave\\ for Spatially Coupled Systems with \\ Applications to LDPC Codes}
%
%
%

\author{Rafah~El-Khatib,~\IEEEmembership{Student Member,~IEEE}
        and~Nicolas~Macris,~\IEEEmembership{Member,~IEEE}
\thanks{R. El-Khatib and Nicolas Macris are with the School
of Computer and Communication Sciences, \'Ecole Polytechnique F\'ed\'erale de Lausanne (EPFL), Lausanne, Switzerland, e-mails: \{rafah.el-khatib,nicolas.macris\}\@epfl.ch.}}

\maketitle

\begin{abstract}
We consider the dynamics of message passing for spatially coupled codes and, in particular,
the set of density evolution equations that tracks the profile of decoding errors along the spatial direction of coupling. 
It is known that, for suitable boundary conditions and after a transient phase,
the error profile exhibits a ``solitonic behavior". Namely, a uniquely-shaped
wavelike solution develops, that propagates with constant velocity. Under this assumption we derive an analytical 
formula for the velocity in the framework of a continuum limit of the spatially coupled system. The general 
formalism is developed for spatially coupled low-density parity-check codes on general binary memoryless symmetric 
channels which form the main system of interest in this work. We apply the formula for special channels and illustrate  that 
it matches the direct numerical evaluation of the velocity for a wide range of noise values. A possible application of 
the velocity formula to the evaluation of finite size scaling law parameters is also discussed. We conduct a similar 
analysis for general scalar systems and illustrate the findings with applications to compressive sensing and generalized 
low-density parity-check codes on the binary erasure or binary symmetric channels. 
\end{abstract}


%

\IEEEpeerreviewmaketitle

\section{Introduction}

\IEEEPARstart{S}{patial} coupling was first introduced by Felstrom and Zigangirov in the context of low-density parity-check (LDPC) codes \cite{Zigangirov}. Spatially coupled codes have been shown to be capacity-achieving  on binary memoryless symmetric (BMS) channels under belief propagation (BP) decoding. 
The capacity-achieving property is due to the ``threshold saturation'' of the BP threshold of the coupled system towards the maximum a-posteriori (MAP) threshold of the 
uncoupled code ensemble \cite{KRUUniv}, \cite{kumar2014threshold}. 
Spatial coupling has also been applied to several other problems
besides channel coding \cite{lentmaier2005terminated}, \cite{lentmaier2010iterative}, 
such as lossy source compression \cite{Aref-Macris-Vuffray-1}, \cite{Aref-Macris-Vuffray-2}, compressive sensing \cite{kudekar2010effect}, \cite{krzakala2012probabilistic}, \cite{donoho2012information}, 
random constraint satisfaction problems \cite{hassani2010itw}, \cite{hassani2013threshold}, \cite{hassanisoda16},
and a coupled Curie-Weiss (toy) model \cite{hassani2012meanfield}, \cite{caltagirone2014dynamics}.



Consider a single (uncoupled) LDPC code. To construct the corresponding coupled code of spatial length $\len+w$, we take $\len+w$ replicas of the single system and ``couple" every $w$ adjacent single systems by means of a uniform window function. At every iteration of the BP algorithm, the variable and check nodes of the coupled graph exchange messages which are described by a set of coupled density evolution (DE) iterative equations. The solution to the DE equations, called the \emph{decoding profile} $\mathtt{x}$, is a vector of ``error distributions" (more precisely, distributions of the BP log-likelihood estimates) along the spatial axis of coupling. More specifically, let the integer $z\in \{-w+1, \dots, \len\}$ denote the position along the spatial direction of the graph construction (on which the replicas are spread). Then the $z^{th}$ component of $\mathtt{x}$, call it $\mathtt{x}_z$, denotes the distribution of the BP log-likelihood estimate at the $z^{\text{th}}$ position. In the special case of the binary erasure channel (BEC) this component is reduced to the usual scalar erasure probability 
$0\leq x_z \leq 1$ at position $z$ along the spatial axis of coupling.

Spatially coupled codes perform well, and are capacity achieving, due to the threshold saturation phenomenon that is proved for general BMS channels in \cite{KRUUniv}, \cite{kumar2014threshold}. More specifically, as long as the channel noise is below the MAP threshold, 
the decoding profile 
of a \emph{spatially coupled} code converges to 
the all-$\Delta_\infty$ vector after enough iterations of the BP algorithm, where $\Delta_\infty$ is the Dirac mass at infinite log-likelihood (i.e. perfect knowledge of the bits). In the special case of the BEC, the all-$\Delta_\infty$ vector corresponds to a vector of scalar erasure probabilities driven to zero by DE iterations. On the other hand, the probability distribution of the log-likelihoods of bits of the corresponding \emph{uncoupled} code only converge to $\Delta_\infty$ when the channel noise is below the BP threshold (which is lower than the MAP threshold).

The threshold saturation phenomenon is made possible due to ``seeding" at the boundaries of the spatially coupled code. Seeding means that we fix the bits at the boundaries so that the probability distributions of their log-likelihoods are $\Delta_\infty$. This facilitates BP decoding near the boundaries, and this effect is propagated along the rest of the coupled chain. The minimum size of the seed that guarantees the propagation of the decoding effect is of the same order as the size $w$ of the coupling window; however, an exact determination of the minimum possible such size is an still an interesting open question.

When the channel noise is between the BP and the MAP thresholds, and the underlying uncoupled ensemble has a 
unique non-trivial stable BP fixed point that blocks decoding, an interesting phenomenon that has been empirically observed 
is the appearance of a \emph{solitonic decoding wave} after a certain number of transient iterations of the BP algorithm. 
This \emph{soliton} is characterized by a \emph{fixed shape} that seems independent of the initial condition and 
has a \emph{constant traveling velocity} that we henceforth call $v$. This phenomenology is discussed 
in more details in Section~\ref{phenomenology}. 
Figures~\ref{fig:transientPhase} and~\ref{fig:wavePropagation} in Section~\ref{phenomenology} show an example of the transient phase and of the soliton 
in the case of spatially coupled codes on the BEC. The main goal of this work is to derive 
a formula for the  velocity of the soliton  for {\it general BMS channels}.

The decoding wave has recently been studied in the context of coding when transmission takes place 
over the BEC. In \cite{KRU12}, it is proved that the solitonic wave solution exists and bounds on 
the velocity of the soliton are derived. However, the independence of the unique shape of the wave 
from the initial conditions remains an open question. In \cite{aref2013convergence}, more 
complex coupled systems are studied, where it is possible to have more than one non-trivial 
stable BP fixed point, and there again some bounds on the velocity of the soliton are provided. The 
solitonic behavior has also been studied for the coupled Curie-Weiss toy model \cite{hassani2012meanfield} and in \cite{caltagirone2014dynamics}
a formula for the velocity, as well as an approximation, are derived and tested numerically.

In the first part of this work, we derive a general formula for the velocity of the wave in the asymptotic
limit $\len\gg w \gg 1$ in the context of coding when transmission 
takes place over {\it general} BMS channels (see Equ. \eqref{eqn:vFormulaBMS}). This limit enables us to formulate the 
problem in a ``continuum limit" which makes the derivations quite tractable. We show, with the use of numerical simulations, that this 
continuum limit yields good approximations for the velocity of the original discrete system. 
For simplicity, we limit ourselves to the case where the underlying uncoupled LDPC code 
has only {\it one non-trivial stable BP fixed point}. 

Our derivation rests on the assumption 
that the soliton indeed appears. More precisely, we assume that after an initial transient phase, 
the decoding profile develops a unique shape, independent of the initial condition, and travels with a constant velocity $v$. 
This assumption can be strictly true 
only in an asymptotic limit of a very large chain length and a 
large iteration number (or time). It is an interesting open problem to make this space-time asymptotic limit precise
and rigorously prove that the soliton appears and is independent of the initial condition. We conjecture that our velocity formula 
is exact in such a limit.

The formula for the velocity of the wave greatly simplifies when we 
consider transmission over the BEC, because the decoding profile reduces to a scalar vector of erasure probabilities. 
For transmission over general BMS channels, we also simplify the analysis by applying the 
Gaussian approximation \cite{chung2001capacity}, \cite{chung2000construction}. This consists of approximating the DE densities and 
the channel by suitable ``symmetric" Gaussian densities. Since the mean $m$ and the variance $\sigma^2$ of these symmetric
Gaussian densities are related by $\sigma^2=2m$, the analysis then reduces to that of a one-dimensional scalar system, whose technical 
difficulty is similar to that of the case of transmission over the BEC. 
We thus obtain a more tractable velocity formula and compare the numerical predictions of these velocity formulas 
with the empirical value of the velocity for finite values of $\len$ and $w$. Good agreement is found, on practically the whole range of values 
within $[\epsilon_{\text{\tiny BP}}, \epsilon_{\text{\tiny MAP}}]$, even for small values of the window size $w$.


It is of theoretical as well as practical interest to have a hold on the analytical expression of the velocity of the wave. 
The velocity is also related to other fundamental quantities that describe a coding system, such as the finite-size scaling law 
that predicts the error probability of finite-length spatially coupled codes. 
In \cite{olmos2014scaling}, the scaling law for a finite-length spatially coupled $(\ell,r,\len)$ code, when transmission takes place over the BEC, is derived. 
Involved in this scaling law are parameters that can be estimated using the value of the velocity of the decoding wave. 
Using values of the velocity computed in our work, we provide reasonably good estimates of these parameters.

In the second part of this work, we consider general spatially coupled scalar bipartite systems 
(that are not restricted to coding) governed by a general message passing algorithm. In this setting, 
the system is scalar (one-dimensional) since the messages exchanged between the nodes are scalar. Due to 
seeding at the boundary, the ``profile" (we no longer call it the ``decoding profile") exhibits the same phenomenology as in coding. Namely, a 
solitonic behavior appears after a short transient phase. We derive a formula for the velocity of the soliton for such systems 
and illustrate it on two applications: compressive sensing and generalized LDPC (GLDPC) codes.

The derivations of the velocity formulas in both parts of the work use the same tools and assumptions. We combine the use of  
the ``{\it potential functional}" introduced and used in 
a series of works \cite{kumar2014threshold}, \cite{hassani2010itw}, \cite{hassani2012meanfield}, \cite{yedla2014simple}, 
\cite{takeuchi2012phenomenological},
as well as the continuum limit $\len\gg w\gg 1$ which makes the derivations analytically tractable. 
The potential is a ``variational formulation" of the message passing algorithm on coding systems. 
It is a functional whose stationary points are the fixed points of the density evolution equations described by this algorithm, and 
has been used to prove threshold saturation in \cite{kumar2014threshold} for general BMS channels. A significant part of the formalism in 
\cite{kumar2014threshold} is used in the present work. 
We also note that potential formulations have been used to characterize the fixed point(s) of general scalar systems at the MAP threshold using displacement 
convexity in \cite{el2013displacement}, \cite{el2014analysis}. An extension of the ideas in these works could shed some light on the question 
of the independence of the soliton's shape from the initial conditions. 


Section~\ref{section:setting} introduces a few preliminary
notions that we will need and reviews the phenomenology of the solitonic wave.
In Section~\ref{mainresultsection}, we formulate the continuum limit 
and state our main formula for the 
velocity of the soliton on general BMS channels; 
the derivation is presented in Section~\ref{section:velocityBMS}.
Comparisons with numerical experiments are presented 
in Section~\ref{section:velocityCodingApplications}. These concern transmission over the
 BEC  as well as general BMS channels in the so-called Gaussian approximation. 
We also discuss a  possible 
application of our formula to scaling laws for finite-size ensembles in Section~\ref{section:scalingLaw}. 
The case of
general scalar spatially coupled systems is treated in Section~\ref{section:velocityScalar}, 
and illustrated for the examples of generalized LDPC codes (on the BEC or BSC channels) and 
compressive sensing. 
We present concluding remarks and propose further directions in 
Section~\ref{section:conclusion}.

A summary of this work has appeared in \cite{rafah-nicolas-ISIT2016}, \cite{elkhatibITW2016}. 

\section{Preliminaries}\label{section:setting}

We consider (almost) the same setting as in \cite{kumar2014threshold} and adopt most of
the notation introduced in that work. 
For more information about the formalism in these preliminaries, one can consult \cite{richardson2008modern}.

We denote by $M(\bar{\mathbb{R}})$ the space of probability measures 
$\mathtt{x}$ on the extended real numbers $\alpha \in \bar{\mathbb{R}} =\mathbb{R}\cup \{\infty\}$. Here
$\alpha\in\bar{\mathbb{R}}$ should be interpreted as a ``log-likelihood variable".
We call the measure $\mathtt{x}$ {\it symmetric} if
$\int_E \mathrm{d}\mathtt{x}(\alpha)=\int_{-E}e^{-\alpha}  \mathrm{d}\mathtt{x}(\alpha)$
for all measurable sets $E\subset\bar{\mathbb{R}}$.

We define an {\it entropy functional} $H:\mathcal{M}\rightarrow\mathbb{R}$ that maps a finite probability measure 
from $M(\bar{\mathbb{R}})$ to a real number. It is defined as 
\begin{align}\label{eqn:entropy}
H(\mathtt{x})=\int\mathrm{d}\mathtt{x}(\alpha)\log_2(1+e^{-\alpha}).
\end{align}
Note that this is a linear functional. Linearity is used in an important way to compute the entropy of
 convex combinations of measures (which also yield  probability
measures). But we will also compute the ``entropy" associated to differences of measures by setting $H(\mathtt{x}_1-\mathtt{x}_2) \equiv H(\mathtt{x}_1) - H(\mathtt{x}_2)$.  
In other words, the entropy functional is extended in an obvious way to the space of signed measures.

In the remainder of this work, we will use the Dirac masses $\Delta_0(\alpha)$ and $\Delta_\infty(\alpha)$ at zero and infinite log-likelihood, that have entropies 
$H(\Delta_0)= 1$ and $H(\Delta_{\infty})=0$,
respectively. 

We will also use the standard variable-node and check-node convolution operators $\varoast$ and $\boxast$ for 
log-likelihood ratio message distributions involved in DE equations \cite{richardson2008modern}. 
For $\mathtt{x}_1$, $\mathtt{x}_2 \in M(\bar{\mathbb{R}})$, the usual convolution $\mathtt{x}_1\varoast\mathtt{x}_2$ 
is the density of 
$$
\alpha = \alpha_1 +\alpha_2,
$$
and $\mathtt{x}_1\boxast\mathtt{x}_2$ is the 
density of 
$$
\alpha=2\tanh^{-1}(\tanh\frac{\alpha_1}{2}\tanh\frac{\alpha_2}{2}).
$$
More formally, for
any measurable set $E\in\mathbb{R}$, the operators are defined by
\begin{align}
\begin{cases}
(\mathtt{x}_1\varoast\mathtt{x}_2)(E)&=\int\mathrm{d}\mathtt{x}_2(\alpha)\mathtt{x}_1(E-\alpha),\\
(\mathtt{x}_1\boxast\mathtt{x}_2)(E)&=\int\mathrm{d}\mathtt{x}_2(\alpha)\mathtt{x}_1\Big(2\tanh^{-1}\Big(\frac{\tanh(E/2)}{\tanh(\alpha/2)} \Big)\Big).
\end{cases}
\end{align}
We note that the identities of the $\varoast$ and $\boxast$ operators are $\Delta_\infty$ and $\Delta_0$, and their 
annihilators are $\Delta_0$ and $\Delta_\infty$. More explicitly,
\begin{align}\label{identityannihilation}
\begin{cases}
\Delta_\infty\boxast\mathtt{x}=\mathtt{x},\quad\Delta_0\boxast\mathtt{x}=\Delta_0,\\
\Delta_0\varoast\mathtt{x}=\mathtt{x},\quad\Delta_\infty\varoast\mathtt{x}=\Delta_\infty.
\end{cases}
\end{align}
Each operation, taken separately, is associative, commutative, and linear. 
However, when they are taken together, there is no distributive law; also, they don't associate in the sense that 
 $\mathtt{x}_1\varoast(\mathtt{x}_2\boxast\mathtt{x}_3)\neq (\mathtt{x}_1\varoast\mathtt{x}_2)\boxast\mathtt{x}_3$ 
 and $\mathtt{x}_1\boxast(\mathtt{x}_2\varoast\mathtt{x}_3)\neq (\mathtt{x}_1\boxast\mathtt{x}_2)\varoast\mathtt{x}_3$.
We will also use the so-called {\it duality rules}
\begin{align}\label{eqn:operatorProp2}
\begin{cases}
H(\mathtt{x}\varoast\mathtt{y})+H(\mathtt{x}\boxast\mathtt{y})=H(\mathtt{x}) +H(\mathtt{y}),\\
H(\mathtt{x}\varoast\mathtt{a})+H(\mathtt{x}\boxast\mathtt{a})=H(\mathtt{a}),\\
H(\mathtt{a}\varoast \mathtt{b})+H(\mathtt{a}\boxast \mathtt{b})=0,\\
\end{cases}
\end{align}
 where $\mathtt{x}, \mathtt{y}\in M(\bar{\mathbb{R}})$ and $\mathtt{a}$, $\mathtt{b}$ are differences 
of probability measures $\mathtt{a}=\mathtt{x}_1-\mathtt{x}_2$, $\mathtt{b}= \mathtt{x}_3-\mathtt{x}_4$, $\mathtt{x}_i\in M(\bar{\mathbb{R}})$, $i=1,2,3,4$.

\subsection{Single System}\label{ssection:singleSystem}

Consider an (uncoupled) LDPC($\lambda,\rho$) code ensemble and transmission over the BMS channel. 
Here, $\lambda(y)=\sum_{l} \lambda_l y^{l-1}$ and $\rho(y)=\sum_{r} \rho_r y^{r-1}$ are the usual edge-perspective variable-node and check-node degree distributions, respectively. The node-perspective degree distributions $L$ and $R$ are  
defined by $L'(y)=L'(1)\lambda(y)$ and $R'(y)=R'(1)\rho(y)$, respectively.
Moreover, consider communication over a family of BMS channels whose 
distribution $\mathtt{c}_\mathtt{h}(\alpha)$ in the log-likelihood domain is parametrized by the 
channel entropy\footnote{In the literature, this quantity is often denoted by $\mathtt{c}(\mathtt{h})$.} $H(\mathtt{c}_\mathtt{h})=\mathtt{h}$. 

Let $\tilde{\mathtt{x}}^{(t)}$ denote the variable-node output distribution of the BP algorithm at iteration $t\in \mathbb{N}$.
We can track the average behavior of the BP decoder by means of the DE iterative equations that are 
written as a recursion in terms of the variable-node output distribution as follows
\begin{align}\label{eqn:DEuncoupled}
\tilde{\mathtt{x}}^{(t+1)}=\mathtt{c}_\mathtt{h}\varoast\lambda^\varoast(\rho^\boxast(\tilde{\mathtt{x}}^{(t)})),
\end{align}
with initial condition $\tilde{\mathtt{x}}^{(0)}=\Delta_0$ (equivalently, we can take the perhaps more natural 
initial condition $\tilde{\mathtt{x}}^{(0)}=\mathtt{c}_\mathtt{h}$).

There are two thresholds of interest for us. The first one is the algorithmic threshold; it is defined 
for a family of BMS channels whose channel distributions $\mathtt{c}_\mathtt{h}(\alpha):\mathbb{R}\rightarrow M(\bar{\mathbb{R}})$ are ordered by 
degradation and parametrized by their entropy $H(\mathtt{c}_\mathtt{h})=\mathtt{h}$. It is also called the BP threshold of the family and is defined as
$$
\mathtt{h}_{\text{\tiny BP}}=\{\mathtt{h}\in[0,1]:
\, \tilde{\mathtt{x}}=\mathtt{c}_\mathtt{h}\varoast\lambda^\varoast(\rho^\boxast(\tilde{\mathtt{x}}))
\implies \tilde{\mathtt{x}}=\Delta_\infty \}.
$$
The second threshold corresponds to optimal (MAP) decoding
$$
\mathtt{h}_{\text{\tiny MAP}}=\{\mathtt{h}\in[0,1]:
\liminf\limits_{n\rightarrow\infty}\frac{1}{n}\mathbb{E}[H(X^n|Y^n(\mathtt{h}))]>0\},
$$
where $H(X^n|Y^n(\mathtt{h}))$ is the conditional Shannon entropy of the input given by the channel 
observations, and $\mathbb{E}$ is the expectation over the code ensemble.

The potential functional $W_s(\mathtt{x})$, $\mathtt{x}\in\mathcal{M}(\bar{\mathbb{R}})$, of the ``single'' or uncoupled system is
\begin{align}
W_s(\mathtt{x})& =\frac{1}{R'(1)}H(R^\boxast(\mathtt{x}))+H(\rho^\boxast(\mathtt{x}))
-H(\mathtt{x}\boxast\rho^\boxast(\mathtt{x}))-\frac{1}{L'(1)} H(\mathtt{c}\varoast L^\varoast(\rho^\boxast(\mathtt{x}))).
\label{eqn:singlePotBMS}
\end{align}
The fixed point form of the DE equation \eqref{eqn:DEuncoupled} is obtained by setting to zero the functional 
derivative of $W_s(\mathtt{x};\mathtt{c})$ with respect to $\mathtt{x}$. In other words, 
$\mathtt{x}=\mathtt{c}_\mathtt{h}\varoast\lambda^\varoast(\rho^\boxast(\mathtt{x}))$ is equivalent to
\begin{align}
\lim_{\gamma \to 0} \frac{1}{\gamma}(W_s(\mathtt{x}) +\gamma\mathtt{\eta}) - W_s(\mathtt{x}))=0,
\end{align}
where $\mathtt{\eta}$ is a difference of two probability measures (see \cite{kumar2014threshold} for the proof of this statement). The BP and MAP thresholds, $\mathtt{h}_{\text{\tiny BP}}$ and $\mathtt{h}_{\text{\tiny MAP}}$, respectively, can be obtained 
from the analysis of the stationary points of the potential function. See \cite{kumar2014threshold}, \cite{Andrei-Ruediger-N} for more 
details and a rigorous discussion of this issue. 

\vskip 0.25cm 
\noindent{\it Remark about notation.}
In the remainder of this work, most of the time, we omit the subscript $\mathtt{h}$ from $\mathtt{c}_{\mathtt{h}}$ and the 
argument $\alpha$ from $\mathtt{x}(\alpha)$. This is because we will need 
a subscript (resp. an argument) $z$ that represents the position along the chain in the discrete (resp. continuous) case.

\subsection{Spatially Coupled System}\label{ssection:SCsystem}

For standard LDPC codes, the BP threshold $\mathtt{h}_{\text{\tiny BP}}$ is, in general, lower than the MAP threshold $\mathtt{h}_{\text{\tiny MAP}}$. The definitions of the BP and MAP thresholds above extend to the spatially coupled setting. Spatial coupling exhibits two attractive properties. First, {\it the MAP threshold is conserved} under coupling
in the limit $\len\to +\infty$ and for all $w$. The proof of this statement is found in \cite{Andrei-Ruediger-N} (see also \cite{ARN-2012}, \cite{ARN-2013}).  
Second, {\it the  BP threshold of the coupled system 
saturates to its MAP threshold} as proved in \cite{KRUUniv}, \cite{kumar2014threshold}.  The main consequence 
of threshold saturation
is that one can 
decode perfectly up to the $\mathtt{h}_{\text{\tiny MAP}}$. 

Let us now describe the density evolution and potential functional formalism for the spatially coupled
code ensemble. Consider $\len+w$ ``replicas" of the single 
system described in Section~\ref{ssection:singleSystem}, placed on the spatial coordinates $z\in\{-w+1,\dots,\len\}$. The system 
at position $z$ is coupled to $w$ neighboring systems by means of a coupling 
window. For simplicity, we consider a uniform coupling window.
We denote by $\tilde{\mathtt{x}}_z^{(t)}$ the variable-node output distribution at position $z\in\{-w+1,\dots,\len\}$ on the spatial axis
and at time $t\in\mathbb{N}$. The DE equation of the coupled system takes the form 
\begin{align}
\tilde{\mathtt{x}}_z^{(t+1)}=\mathtt{c}_{z}
\varoast\lambda^\varoast\Bigg(\frac{1}{w}\sum\limits_{i=0}^{w-1}\rho^\boxast\Big(\frac{1}{w}\sum\limits_{j=0}^{w-1}\tilde{\mathtt{x}}_{z+i-j}^{(t)} \Big) \Bigg).
\label{eqn:DEcoupledDiscreteVNO}
\end{align}
In this equation, $\mathtt{c}_z=\mathtt{c}$, for $z\in\{1,\dots, \len\}$ and $\mathtt{c}_z=\Delta_\infty$ for 
$z\in \{-w+1, \dots, 0\}$. Furthermore, we fix the left boundary to $\mathtt{x}_z^{(t)}=\Delta_\infty$  for $z\in\{-w+1,\dots,0\}$, for all $t\in\mathbb{N}$. These conditions express perfect 
information at the left boundary which is what enables seeding and the decoding wave propagation along the chain of coupled codes. The initial condition \eqref{eqn:DEcoupledDiscreteVNO} is $\mathtt{x}_z^{(0)} = \Delta_0$ for 
$z\in \{1, \dots, \len+w-1\}$. 

It will be convenient to work with a smoothed version of the profile $\tilde{\mathtt{x}}_z^{(t)}$, namely
 $\mathtt{x}_z^{(t)}=\frac{1}{w}\sum\limits_{i=0}^{w-1}\tilde{\mathtt{x}}_{z-i}^{(t)}$, which is the check-node input distribution. Then, using this change of variables, \eqref{eqn:DEcoupledDiscreteVNO} can be rewritten as
\begin{align}
\mathtt{x}_z^{(t+1)}=\frac{1}{w}\sum\limits_{i=0}^{w-1}\mathtt{c}_{z-i}
\varoast\lambda^\varoast\Bigg(\frac{1}{w}\sum\limits_{j=0}^{w-1}\rho^\boxast\Big(\mathtt{x}_{z-i+j}^{(t)} \Big) \Bigg).
\label{eqn:DEcoupledDiscrete}
\end{align}
Just as in the single system case, this DE equation can be expressed as the stationarity condition of a potential functional (see \cite{kumar2014threshold})
\begin{align}
W(\underline{\mathtt{x}})= & \sum\limits_{z=-w+1}^{L}\Big\{\frac{1}{R'(1)}H(R^\boxast(\mathtt{x}_z))
+H(\rho^\boxast(\mathtt{x}_z)) -H(\mathtt{x}_z\boxast\rho^\boxast(\mathtt{x}_z))
-\frac{1}{L'(1)} H\Big(\mathtt{c}_z\varoast L^\varoast\big(\frac{1}{w}\sum\limits_{i=0}^{w-1}\rho^\boxast(\mathtt{x}_{z+i})\big)\Big)\Big\}.
\label{coupledpotentialdiscrete}
\end{align}
where $\underline{\mathtt{x}} = (\mathtt{x}_{-w+1}, \dots, \mathtt{x}_{L+w-1})$.
The fixed point form of  \eqref{eqn:DEcoupledDiscrete} is equivalent to 
$\lim_{\gamma\to 0}\gamma^{-1}(W(\underline{\mathtt{x}}+\gamma\underline{\mathtt{\eta}}) - W(\underline{\mathtt{x}}))=0$ for 
$\underline{\mathtt{\eta}}=(\mathtt{\eta}_{-w+1}, \dots, \mathtt{\eta}_{L+w-1})$ where $\mathtt{\eta}_i$ are differences of probability measures. 

\subsection{Phenomenological observations}\label{phenomenology}

Our derivation is far from rigorous and is based on an assumption derived from a 
phenomenological picture observed from simulations. We summarize the main observations in this paragraph for the case of transmission over the BEC channel. This channel also gives us the opportunity to illustrate 
the formalism outlined in Sections~\ref{ssection:singleSystem} and~\ref{ssection:SCsystem} in a concrete case.

\begin{figure}
\centering
\includegraphics[draft=false,scale=0.27]{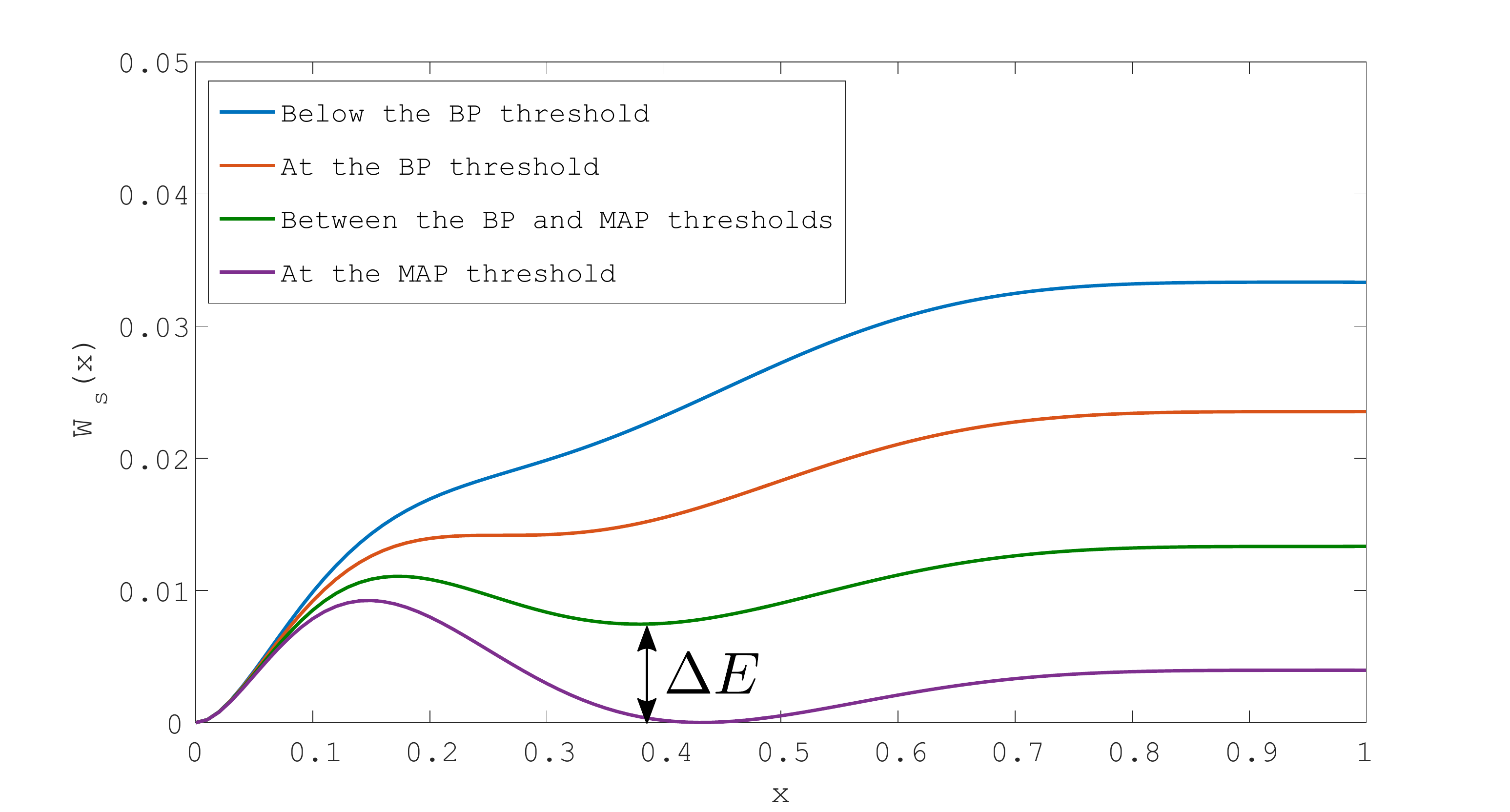}
\caption{We plot the potential function of the $(3,6)$ regular LDPC when transmission takes place over the BEC$(\epsilon)$ for several values of 
the erasure probability $\epsilon$. Note that $x=0$ is always a trivial stationary point. For $\epsilon<0.4294 =\epsilon_{\text{\tiny BP}}$ the potential function is strictly increasing and $x=0$ is the only minimum. At $\epsilon_{\text{\tiny BP}}=0.4294$ a horizontal inflexion point develops and a second non-trivial minimum 
exists for larger $\epsilon$.  At $\epsilon_{\text{\tiny MAP}} = 0.4881$ the trivial and non-trivial minimum are at the same height and the energy gap $\Delta E$ vanishes.}
\label{fig:potentialSeveralEps}
\end{figure}

The BEC has channel distribution $\mathtt{c}_\epsilon(\alpha)=\epsilon\Delta_0+(1-\epsilon)\Delta_\infty$, where 
$\epsilon$ is the erasure probability, and $H(\mathtt{c}_\epsilon)=\epsilon$ (hence $\mathtt{h}=\epsilon$).
The density of the BP estimates of log-likelihood variables can be parametrized as 
$\mathtt{x}^{(t)}(\alpha) = x^{(t)}\Delta_0(\alpha) + (1-x^{(t)})\Delta_{\infty}(\alpha)$, 
where $x^{(t)}\in[0,1]$ is interpreted as the erasure probability at iteration $t\in\mathbb{N}$.
The DE equation becomes a one-dimensional iterative map 
\begin{align}
x^{(t+1)} = \epsilon\lambda(1-\rho(1-x^{(t)}))
\end{align}
over scalars in $[0, 1]$. These iterations are always initialized with $x^{(0)}=1$ or, equivalently, $x^{(0)} = \epsilon$.  The corresponding fixed point equation is the stationarity condition for the potential 
function
\begin{align}
W_{\text{\tiny BEC}}(x) & =\frac{1}{R'(1)}(1-R(1-x))-x\rho(1-x)-\frac{\epsilon}{L'(1)}L(1-\rho(1-x)).
\label{potential-BEC-first-time}
\end{align}
Note that the potential function is defined up to a constant which is set here so that $W_{\text{\tiny BEC}}(0) =0$.
Figure~\ref{fig:potentialSeveralEps} illustrates  the potential function
for a $(3,6)$-regular Gallager ensemble, for several values of $\epsilon$.  
For $\epsilon<0.4294$, the potential function \eqref{potential-BEC-first-time} is strictly increasing, and equivalently the DE iterations
are driven to the unique minimum at $x=0$. At $\epsilon_{\text{\tiny BP}} = 0.4294$ a horizontal inflexion point appears and a second 
non-trivial local minimum $x_{\text{\tiny BP}}$ appears; this minimum corresponds to the non-trivial fixed point reached by DE iterations. It is known that the MAP threshold is equal to the erasure probability where the 
non-trivial local minimum is at the same height as the trivial one and that decoding becomes impossible once the non-trivial minimum becomes a global minimum. For this example, this happens when $\epsilon_{\text{\tiny MAP}} = 0.4881$. 
Figure~\ref{fig:potentialSeveralEps} also shows the {\it energy gap} that is defined for $\epsilon_{\text{\tiny BP}}\leq\epsilon\leq\epsilon_{\text{\tiny MAP}}$ as $\Delta E= W_{\text{\tiny BEC}}(x_{\text{\tiny BP}}) - W_{\text{\tiny BEC}}(0)$. At the MAP threshold, 
we have $\Delta E =0$. 

Let us now describe the phenomenology of the soliton (decoding wave) for spatially coupled codes. Our discussion is limited to 
the case where the underlying code ensemble has a single non-trivial DE fixed point (equivalently, the potential function has a single non-trivial local minimum). One can show that this is always the case for regular code ensembles. For irregular degree distributions the situation may be more complicated with many non-trivial fixed points that appear. For transmission over the BEC, equation \eqref{eqn:DEcoupledDiscrete} reads 
\begin{align}
x_z^{(t+1)}=\frac{1}{w}\sum\limits_{i=0}^{w-1}\epsilon_{z-i}
\lambda\Big(\frac{1}{w}\sum\limits_{j=0}^{w-1}\big(1- \rho(1- x_{z-i+j}^{(t)})\big) \Big).
\end{align}
Here, $\epsilon_z=0$ for 
$z\in \{-w+1, \dots, 0\}$ and $\epsilon_z=\epsilon$ for $z\in\{1,\dots,\len\}$. Furthermore, we fix the left boundary to $x_z^{(t)}=0$  for $z\in\{-w+1,\dots,0\}$, for all $t\in\mathbb{N}$. These are the ``perfect seeding'' conditions which enable the initiation of decoding. The initial condition for the 
iterations is $x_z^{(0)} = 1$ (or $\epsilon$) for 
$z\in \{1, \dots, \len+w-1\}$. 

\begin{figure}
\centering
\includegraphics[draft=false,scale=0.27]{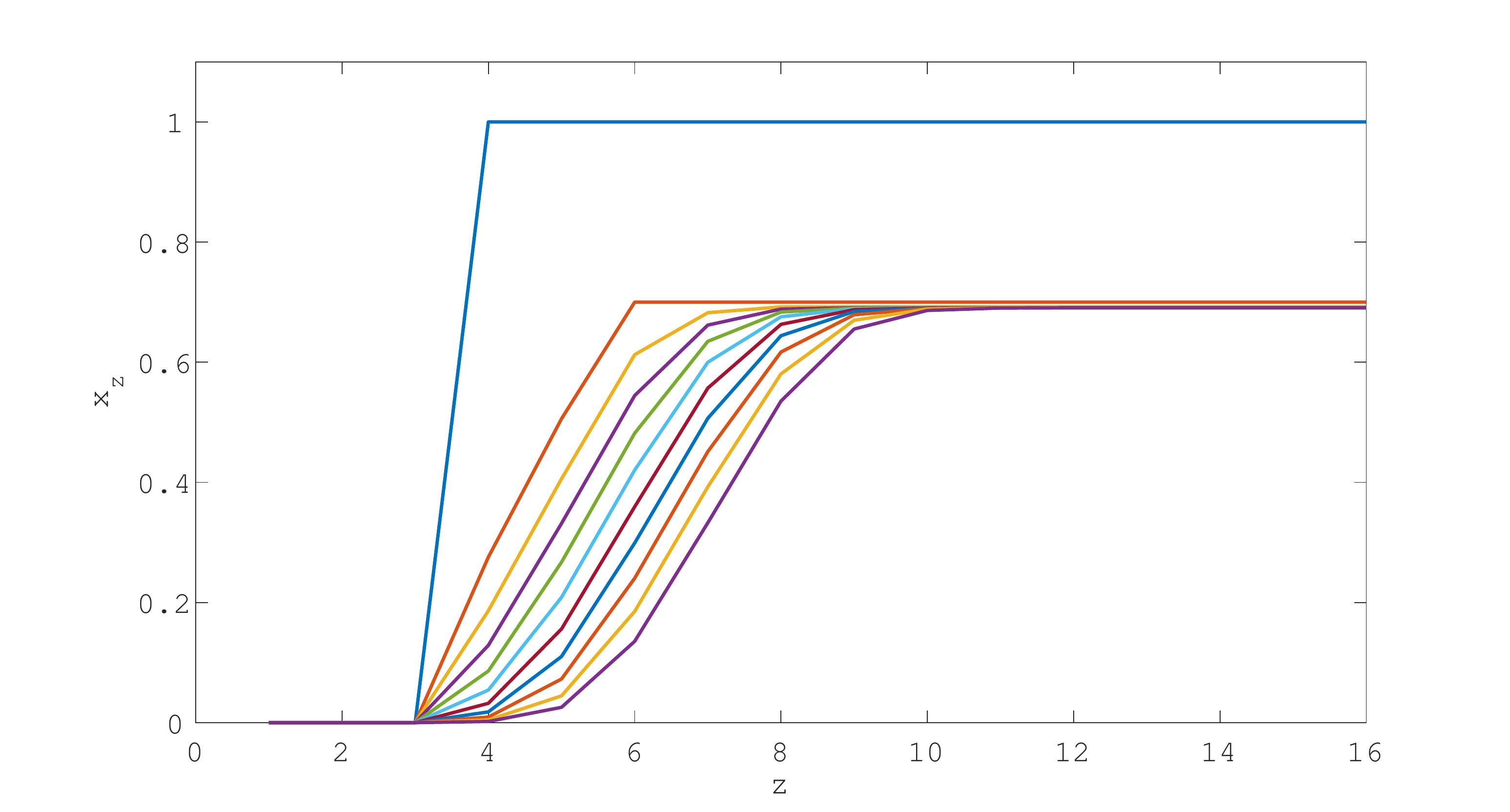}
\caption{We consider the spatially coupled LDPC($\lambda(x),\rho(x)$) ensemble, with $\lambda(x)=0.3x^3+0.4x^5+0.3x^6$ and $\rho(x)=x^5$, with transmission over the BEC($0.7$). The parameters of coupling are $\len+w=16$ and $w=3$. We plot the decoding profile during the first 10 iterations, where the profile is initialized to 0 when $z\leq 3$ and to 1 elsewhere. We observe that the segment initialized to 1 decreases quickly and converges to the BP threshold $x_{\text{\tiny BP}}=0.6907$. We observe that the transient phase is only about 5 iterations here, before the decoding profile converges to a fixed shape.}
\label{fig:transientPhase}
\end{figure}

\begin{figure}
\centering
\includegraphics[draft=false,scale=0.27]{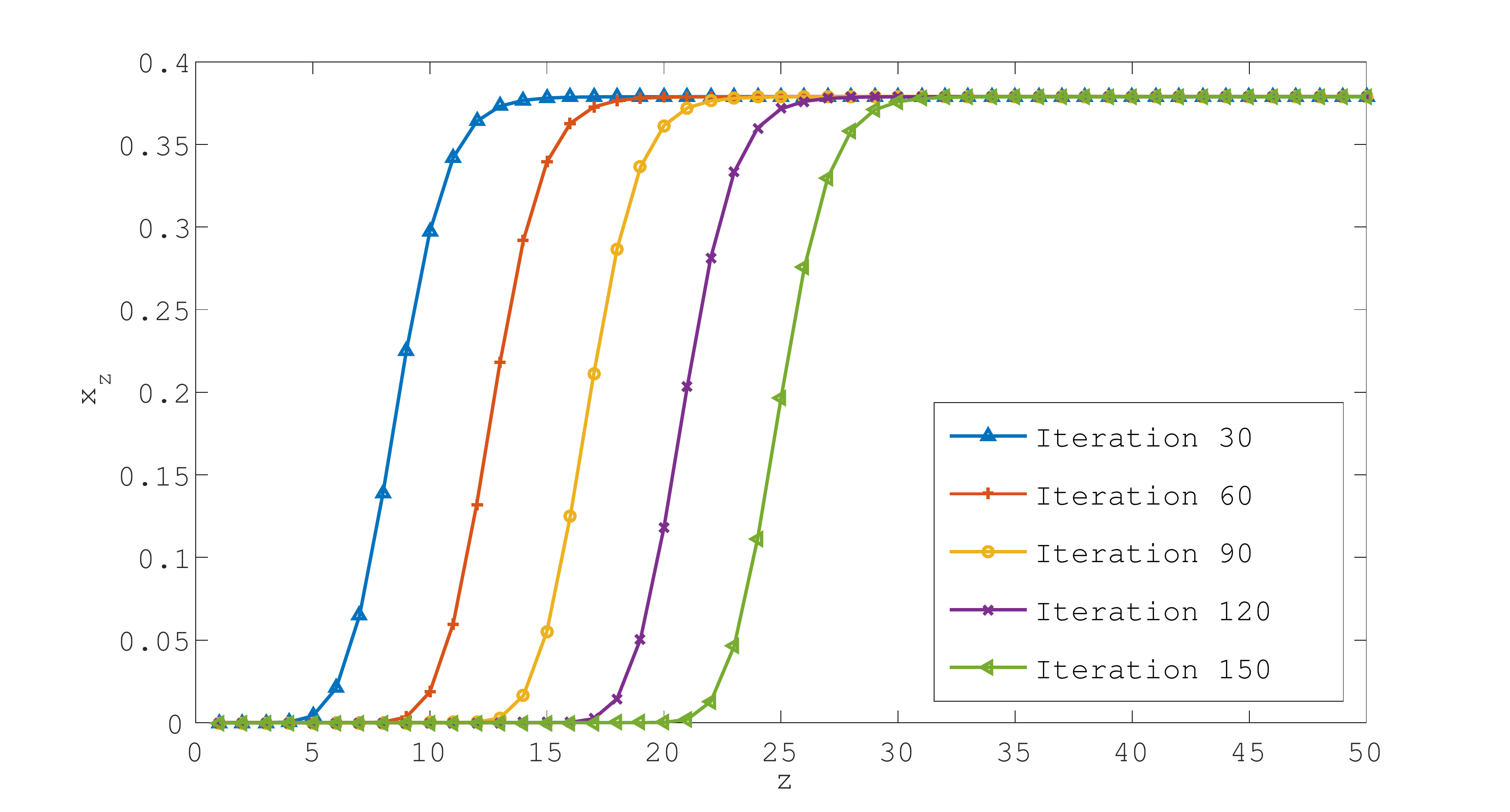}
\caption{We consider the $(3,6)$-regular LDPC spatially coupled code with $L=50$, $w=3$ on the BEC($0.46$). 
We plot the error probability along the spatial dimension and observe the ``decoding wave". This ``soliton" 
is plotted every 30 iterations until iteration 150 and is seen to make a quick transition from zero error 
probability to the BP-value $x_{\text{\tiny BP}}=0.3798$ of the error probability. The optimal (MAP) noise threshold is 
$\epsilon_{\text{\tiny MAP}} = 0.4881.$}
\label{fig:wavePropagation}
\end{figure}

The evolution of the decoding wave can be decomposed into two phases: a {\it transient} and a {\it stationary} phase.
In the transient phase, we observe a profile of erasure probabilities $x_z$ changing shape. The segment initialized to $x_z^{(0)} = 1$ quickly drops to $x_z \approx x_{\text{\tiny BP}}$ where it remains stuck on the far right for large values of $z$. The seeding region, on the other hand, starts progressing towards the right-hand side and, after a few iterations, a fixed profile shape develops. This transient phase is illustrated in Figure~\ref{fig:transientPhase}
for an irregular code. Overall, it only lasts for a few iterations (of the order of $5$ iterations in this example).  
After this transient phase is over, one observes a stationary phase with a {\it solitonic  behavior}, as depicted in Figure~\ref{fig:wavePropagation}. 
The profile of erasure probabilities has a stationary shape with a front at position $z_{\text{\tiny front}}$ that moves, at a  constant speed, towards the right. The soliton is relatively well-localized within approximately $2w$ positions 
 and quickly approaches $x_z \to 0$ for $z < z_{\text{\tiny front}}$ and $x_z \to x_{\text{\tiny BP}}$ 
for $z > z_{\text{\tiny front}}$. The stationary phase and its soliton are depicted in 
Figure~\ref{fig:wavePropagation} for a finite spatially coupled $(3,6)$-regular ensemble with chain length $\len=50$ and $w=3$ for $\epsilon =0.46$. In this figure, we plot the decoding profile every 30 iterations starting at the $30^{th}$ iteration (the leftmost curve) and until the $150^{th}$ iteration (the rightmost curve). The kink increases sharply from $x_z =0$ to $x_z=x_{\text{\tiny BP}}= 0.3789$ over a width of the order of $2w =6$. 

\section{Continuum limit and main result}\label{mainresultsection}

\subsection{Continuum Limit}\label{ssection:continuumLimit}

We now consider the coupled system in the {\it continuum limit}, in which the length of the coupling chain $\len$ is first taken very large $\len\rightarrow +\infty$, and then the window size is taken very large $w\rightarrow +\infty$. The continuum limit has already been considered for the special case of the BEC in \cite{KRU12}, \cite{el2013displacement}, \cite{el2014analysis}.
We slightly abuse notation by keeping the same symbols for the profile, the spatial position, and the channel distribution in the continuum limit. Thus, we denote by $\mathtt{x}(\cdot,\cdot)$ the \emph{continuous} profile of distributions and set $\mathtt{x}(\frac{z}{w},t)\equiv\mathtt{x}_z^{(t)}$. We then replace $\frac{z}{w}\to z$ so that the new $z$ is the \emph{continuous} variable on the spatial axis, $z\in\mathbb{R}$.  

In view of the discussion of the phenomenology in Section~\ref{phenomenology}, we consider the class of profiles satisfying the ``natural boundary conditions" $\mathtt{x}(z,t)\to\Delta_\infty$ when $z\to - \infty$ for all $t\in\mathbb{R}$,  $\mathtt{x}(z,t)\to\mathtt{x}_{\text{\tiny BP}}$ when $z\to+\infty$ for all $t\in\mathbb{N}$, where  $\mathtt{x}_{\text{\tiny BP}}$ is the unique non-trivial stable fixed point of DE for the single system Equ. \eqref{eqn:DEuncoupled}. 

The BMS channel distribution is now also continuous, and we denote by $\mathtt{c}(z)$ the channel distribution at the continuous spatial position $z\in\mathbb{R}$. 
The DE equation \eqref{eqn:DEcoupledDiscrete} then takes the form
\begin{align}
\mathtt{x}&(z,t+1) 
=\int_0^1\mathrm{d}u\,\mathtt{c}(z-u)\varoast\lambda^\varoast\Big(\int_0^1\mathrm{d}s\,\rho^\boxast\big(\mathtt{x}(z-u+s,t) \big) \Big).
\label{continuous-DE}
\end{align}
The initial condition at $t=0$  is given by a profile $\mathtt{x}(z, 0)$ that interpolates between the two limiting values 
of the boundary condition, namely $\mathtt{x}(z,0)\to\Delta_\infty$ when $z\to - \infty$ 
and $\mathtt{x}(z,0)\to\mathtt{x}_{\text{\tiny BP}}$ when $z\to+\infty$.

\subsection{Statement of Main Result}\label{ssection:vBMSstatement}

We consider the channel entropy $\mathtt{h}\in [\mathtt{h}_{\text{\tiny BP}}, \mathtt{h}_{\text{\tiny MAP}}]$.
The phenomenology tells us that: (i) after a transient phase, the 
profile develops a fixed shape
$\mathtt{X}(\cdot)$; 
(ii) the shape is independent of the initial condition; (iii) the shape travels at constant 
speed $v$; (iv) the shape satisfies the boundary 
conditions $\mathtt{X}(z)\to \Delta_{\infty}$ for $z\to -\infty$ and $\mathtt{X}(z)\to \mathtt{x}_{\text{\tiny BP}}$ for $z\to +\infty$. 
We thus formalize these observations and make an ansatz:

\vskip 0.25cm 
\noindent{\bf Ansatz.} For each $\mathtt{h}\in [\mathtt{h}_{\text{\tiny BP}}, \mathtt{h}_{\text{\tiny MAP}}]$ there exist a constant $v\geq 0$ and a family of probability measures  $\mathtt{X}(z)$ (indexed by $z\in \mathbb{R}$) satisfying the boundary conditions
$\mathtt{X}(z)\to \Delta_{\infty}$ for $z\to -\infty$ and $\mathtt{X}(z)\to \mathtt{x}_{\text{\tiny BP}}$ for $z\to +\infty$,
such that, for 
$t\to +\infty$ and $\vert z-vt\vert =O(1)$, the solution of DE \eqref{continuous-DE} is independent of the initial condition and satisfies  $\mathtt{x}(z,t) \to \mathtt{X}(z-vt)$.
\vskip 0.25cm

Implicit in this ansatz is that we restrict ourselves here to underlying code ensembles that have only one non-trivial (stable) BP fixed point.
This is true for regular codes for example (but is not limited to this case). Ensembles with many non-trivial fixed points 
could lead to more complicated phenomenologies as emphasized in \cite{aref2013convergence} and would require to modify
the ansatz.

\vskip 0.25cm 
\noindent{\bf Velocity of the soliton for general BMS channels.}
Under the assumption above the velocity of the soliton is given by
\begin{align}\label{eqn:vFormulaBMS}
v=\frac{\Delta E}{\int_{\mathbb{R}}\mathrm{d}z\,H\Big(\rho^{\prime\boxast}(\mathtt{X}(z))\boxast\mathtt{X}'(z)^{\boxast 2}\Big)},
\end{align}
where $\Delta E$ is the {\it energy gap} defined as
\begin{align}\label{energygap}
\Delta E=W_s(\mathtt{x}_{\text{\tiny BP}})-W_s(\Delta_\infty),
\end{align}
and we recall that $W_s$ is the potential \eqref{eqn:singlePotBMS} of the {\it uncoupled system} , $\mathtt{x}_{\text{\tiny BP}}$ 
is the non-trivial BP fixed point to which 
the uncoupled system converges, and $\Delta_\infty$ is the trivial fixed point (Dirac mass at infinity).
\vskip 0.25cm 

Let us make a few remarks. 
In this formula,
the prime denotes the derivative
$\mathtt{X}^\prime(z) = \lim_{\delta\to 0}\delta^{-1}(\mathtt{X}(z+\delta) -\mathtt{X}(z))$
which is to be interpreted as a difference between two measures.
The energy gap is only defined for $\mathtt{h}_{\text{\tiny BP}}\leq\mathtt{h}\leq\mathtt{h}_{\text{\tiny MAP}}$, 
that is, when the single potential $W_s$ has a non-trivial 
non-negative local minimum (see e.g. Figure~\ref{fig:potentialSeveralEps}). 
It is exactly equal to zero when $\mathtt{h}=\mathtt{h}_{\text{\tiny MAP}}$, which confirms the 
fact that the velocity of decoding is zero (no decoding occurs) in this case. Note also 
that, with our normalizations, $W_s(\Delta_{\infty})=0$.

Formula \eqref{eqn:vFormulaBMS} involves the shape $\mathtt{X}(\cdot)$. Using the DE equation, the ansatz $\mathtt{x}(z,t) \to  \mathtt{X}(z-vt)$, and the approximation 
$\mathtt{x}(z,t+1)-\mathtt{x}(z,t)\approx -v\mathtt{X}'(z-vt)$, valid for small $v$, we find after a change of variables that $\mathtt{X}(z)$ is the solution of 
\begin{align}
\mathtt{X}(z)&-v\mathtt{X}'(z)
=\int_0^1\mathrm{d}u\,\mathtt{c}(z-u)\varoast\lambda^\varoast\Big(\int_0^1\mathrm{d}s\,\rho^\boxast\Big(\mathtt{X}(z-u+s) \Big) \Big).
\label{eqn:diffEqnXShape}
\end{align}
To obtain the shape $\mathtt{X}(z)$ and the velocity $v$, one must iteratively solve the closed system of equations formed by \eqref{eqn:vFormulaBMS} and \eqref{eqn:diffEqnXShape}. Note that the assumption of small $v$ used 
above is strictly valid for $\mathtt{h}$ close to $\mathtt{h}_{\text{\tiny MAP}}$. However, numerical simulations confirm that in practice the resulting velocity formula is precise over the whole range 
$[\mathtt{h}_{\text{\tiny BP}}, \mathtt{h}_{\text{\tiny MAP}}]$.

\section{Derivation of velocity formula for BMS channels}\label{section:velocityBMS}

Let us briefly outline of the main steps of derivation. We first write down a potential functional which 
gives, in the continuous setting, the DE fixed point equation corresponding to \eqref{continuous-DE}. This enables us to formulate 
the DE iterations 
as a sort of gradient descent equation (Section~\ref{gradient-descent}). From there on, we use the ansatz in Section~\ref{ssection:vBMSstatement}
to derive the velocity formula \eqref{eqn:vFormulaBMS}.

\subsection{Density evolution as gradient descent}\label{gradient-descent}

We call $\Delta\mathcal{W}(\mathtt{x})$ the potential functional of the coupled system in the continuum limit obtained from 
\eqref{coupledpotentialdiscrete}. This limit involves an integral over the spatial direction $z\in \mathbb{R}$ and, in order to 
get a convergent result, we must subtract a ``reference energy''. Essentially, any static reference profile, here called $\mathtt{x}_0(z)$, that
satisfies the boundary conditions $\mathtt{x}_0(z)\to \Delta_{\infty}$, $z\to -\infty$ and $\mathtt{x}_0(z)\to x_{\text{\tiny BP}}$, $z\to +\infty$, 
will do the job. For concreteness, one can take a Heaviside-like profile $\mathtt{x}_0(z)= \Delta_{\infty}$, $z<0$, 
$\mathtt{x}_0(z) = x_{\text{\tiny BP}}$, $z\geq 0$.
The potential functional is thus defined as follows, 
\begin{align}
 \Delta\mathcal{W}(\mathtt{x}) = \int_{\mathbb{R}} \mathrm{d}z \, \Big( P(z, \mathtt{x}) - P(z, \mathtt{x_0})\Big),
 \label{diff-P}
\end{align}
where $P(z, \mathtt{x})$ is a $z$-dependent functional of $\mathtt{x}$ equal to 
\begin{align}
P(z, \mathtt{x})
 =
\frac{1}{R'(1)}H(R^\boxast(\mathtt{x}(z,t)))
&+H(\rho^\boxast(\mathtt{x}(z,t)))
 -H(\mathtt{x}(z,t)\boxast\rho^\boxast(\mathtt{x}(z,t)))
\nonumber\\&-\frac{1}{L'(1)}H\Big(\mathtt{c}(z)\varoast L^\varoast\big(\int_0^1\mathrm{d}s\,\rho^\boxast(\mathtt{x}(z+s,t))\big)\Big).
\label{eqn:coupledPotCont}
\end{align}

In Appendix~\ref{appendix}, we calculate the functional derivative of $\Delta\mathcal{W}(\mathtt{x})$
in a direction $\mathtt{\eta}(z, t)$ defined as 
\begin{align}
 \frac{\delta \Delta \mathcal{W}}{\delta \mathtt{x}}[\mathtt{\eta}(z,t)]
 =
 \frac{\mathrm{d}}{\mathrm{d}\gamma}\Delta \mathcal{W}(\mathtt{x}+\gamma \mathtt{\eta})\Big|_{\gamma=0},
\end{align}
and find 
\begin{align}
\frac{\delta \Delta \mathcal{W}}{\delta \mathtt{x}}[\eta(z,t)]
= \int_{\mathbb{R}}\mathrm{d}z\,H\Bigg(\Big(\int_0^1\mathrm{d}u \,& \mathtt{c}(z-u)\varoast\lambda^\varoast \big(\int_0^1\mathrm{d}s \, \rho^\boxast(\mathtt{x}(z-u+s,t)) \big) 
 - \mathtt{x}(z,t)\Big)\boxast\rho^{\prime\boxast}(\mathtt{x}(z,t))\boxast \eta(z,t) \Bigg)
\label{funct-derivative-conti}
\end{align}
From \eqref{continuous-DE} and \eqref{funct-derivative-conti} we deduce that 
\begin{align}
 \int_{\mathbb{R}}\mathrm{d}z\,H\Big((\mathtt{x}(z, t+1) & - \mathtt{x}(z,t))\boxast\rho^{\prime\boxast}(\mathtt{x}(z,t))\boxast \eta(z,t) \Big)
 = \frac{\delta \Delta \mathcal{W}}{\delta \mathtt{x}}[\eta(z,t)]
 \label{eqn:FtnalDeriv2}
\end{align}
We note that, using the duality rule \eqref{eqn:operatorProp2} 
for $\mathtt{a} = \mathtt{x}(z, t+1) - \mathtt{x}(z,t)$ and $b = \rho^{\prime\boxast}(\mathtt{x}(z,t))\boxast \eta(z,t)$ (recall that $\eta$ must be
a difference of
two measures so that $b$ also is such a difference) and the associativity of $\boxast$, this equation can also be formulated as 
\begin{align}
 \int_{\mathbb{R}}\mathrm{d}z\,H\Big((\mathtt{x}(z, t+1) & - \mathtt{x}(z,t))\varoast(\rho^{\prime\boxast}(\mathtt{x}(z,t))\boxast \eta(z,t)) \Big)
 = - \frac{\delta \Delta \mathcal{W}}{\delta \mathtt{x}}[\eta(z,t)]
 \label{eqn:FtnalDeriv3}
\end{align}
In this form, we recognize a sort of infinite-dimensional gradient descent equation in a space of measures. This reformulation 
of DE forms the basis of the derivation of the velocity formula. 

\subsection{Final steps of the derivation}\label{ssection:vBMSderivation}

The potential functional can be decomposed in a ``single system'' part and a contribution that contains 
the ``interaction'' due to coupling. We have
\begin{equation}
 \Delta\mathcal{W}(\mathtt{x}) = \mathcal{W}_s(\mathtt{x}) + \mathcal{W}_i(\mathtt{x}),
\end{equation}
with
\begin{align}
 \mathcal{W}_s(\mathtt{x}) &= \int_{\mathbb{R}} \mathrm{d}z \, \{ P_s(z, \mathtt{x}) - P_s(z, \mathtt{x_0})\},
 \label{diff-Ps}\\
 \mathcal{W}_i(\mathtt{x}) &= \int_{\mathbb{R}} \mathrm{d}z \, \{ P_i(z, \mathtt{x}) - P_i(z, \mathtt{x_0})\},
 \label{diff-Pi}
\end{align}
where 
\begin{align}
P_s(z, \mathtt{x}) =\frac{1}{R'(1)}H(R^\boxast(\mathtt{x}(z,t)&))+H(\rho^\boxast(\mathtt{x}(z,t)))
-H(\mathtt{x}(z,t)\boxast\rho^\boxast(\mathtt{x}(z,t)))
\nonumber \\ &
-\frac{1}{L'(1)}H\big(\mathtt{c}(z)\varoast L^\varoast\big(\rho^\boxast(\mathtt{x}(z,t))\big) \big),
\end{align}
and 
\begin{align}
 P_i(z, \mathtt{x})  = H\Big(\mathtt{c}(z)\varoast L^\varoast\big(\rho^\boxast(\mathtt{x}(z,t))\big)\Big)
-H\Big(\mathtt{c}(z)\varoast L^\varoast\big(\int_0^1\mathrm{d}u\,\rho^\boxast(\mathtt{x}(z+u,t))\big)\Big).
\end{align}
Note, for future use, that in fact $P_s(z, \mathtt{x}) = W_s(\mathtt{x}(z,t))$ is the single system potential \eqref{eqn:singlePotBMS} ``at position $z$''.
With these definitions, the gradient descent equation \eqref{eqn:FtnalDeriv3} can be written as 
\begin{align}
 \int_{\mathbb{R}}\mathrm{d}z\, H\Big( & (\mathtt{x}(z, t+1) - \mathtt{x}(z,t))\boxast\rho^{\prime\boxast}(\mathtt{x}(z,t))\boxast \eta(z,t) \Big)
 = \frac{\delta \mathcal{W}_s}{\delta \mathtt{x}}[\eta(z,t)] + \frac{\delta \mathcal{W}_i}{\delta \mathtt{x}}[\eta(z,t)].
 \label{decomposition-s-i}
\end{align}

Now, we use the ansatz to compute the three terms in this equation in the 
regime $t\to +\infty$, $z\to +\infty$ such that $\vert z - vt\vert = O(1)$. We will choose the 
direction $\eta(z, t) = \mathtt{X}^\prime(z - vt)$.

We start with the left-hand side of \eqref{decomposition-s-i}.
From 
$\mathtt{x}(z,t)\to \mathtt{X}(z-vt)$ and the approximation $\mathtt{x}(z, t+1) - \mathtt{x}(z,t)\approx -v\mathtt{X}'(z-vt)$,
together with the special choice of $\eta(z, t)$,
we can rewrite the left hand side of \eqref{decomposition-s-i}
as
\begin{align}
v\int_{\mathbb{R}}\mathrm{d}z\,H\big(\mathtt{X}'(z-vt)\boxast\rho^{\prime\boxast}(\mathtt{X}(z-vt))\boxast \mathtt{X}^\prime(z,t) \big).
\end{align}
Using the commutativity of the operator $\boxast$, this is equal to 
\begin{align}\label{velo}
v\int_{\mathbb{R}}\mathrm{d}z\,H\big(\rho^{\prime\boxast}(\mathtt{X}(z-vt))\boxast \mathtt{X}'(z-vt)^{\boxast 2} \big).
\end{align}
Note that we can shift the argument in the integrals $z-vt \to z$, and this term becomes independent of time. 

Now, we consider the first functional derivative on the right hand side of \eqref{decomposition-s-i}, when $\eta(z, t) = \mathtt{X}^\prime(z-vt)$.
It should be clear that we can immediately make the change of variables in the integrals $z-vt \to z$ which simplifies the formulas. 
By the calculations in Appendix~\ref{appendix}, we find 
\begin{align}
& \frac{\delta \mathcal{W}_s}{\delta \mathtt{X}}[\mathtt{X}'(z)] =
\int_{\mathbb{R}}\mathrm{d}z\,\biggl\{H\big(\mathtt{X}(z)\varoast[\rho^{\prime\boxast}(\mathtt{X}(z))\boxast \mathtt{X}'(z)] \Big)
-H\big(\mathtt{c}\varoast\lambda^\varoast(\rho^\boxast(\mathtt{X}(z)))
\varoast[\rho^{\prime\boxast}(\mathtt{X}(z))\boxast \mathtt{X}'(z)] \big)\biggr\}.
\label{driva-Ws}
\end{align}
In order to simplify the above, we remark the following
\begin{align*}
&\frac{\mathrm{d}}{\mathrm{d}z}\Big\{\frac{1}{R'(1)}H(R^\boxast(\mathtt{X}(z)))-H(\mathtt{X}(z)\boxast\rho^\boxast(\mathtt{X}(z)))
+H(\rho^\boxast(\mathtt{X}(z)))-\frac{1}{L'(1)}H\big(\mathtt{c}(z)\varoast L^\varoast(\rho^\boxast(\mathtt{X}(z)))\big)\Big\}
\nonumber\\
&
=H(\rho^\boxast(\mathtt{X}(z))\boxast\mathtt{X}'(z))-H(\mathtt{X}'(z)\boxast\rho^\boxast(\mathtt{X}(z)))
-H(\mathtt{X}(z)\boxast\rho^{\prime\boxast}(\mathtt{X}(z))\boxast\mathtt{X}'(z))+H(\rho^{\prime\boxast}(\mathtt{X}(z))\boxast\mathtt{X}'(z))
\nonumber \\
&
\qquad\qquad\qquad\qquad\qquad-H(\mathtt{c}(z)\varoast\lambda^\varoast(\rho^\boxast(\mathtt{X}(z)))\varoast[\rho^{\prime\boxast}(\mathtt{X}(z))\boxast\mathtt{X}'(z)] ).
\end{align*}
Noticing that the first two terms on the right-hand side cancel out, and using the duality rule \eqref{eqn:operatorProp2} for the third term, 
we get the integrand in \eqref{driva-Ws}. In other words, the integrand in \eqref{driva-Ws} equals 
$\frac{\mathrm{d}}{\mathrm{d}z}P_s(z, \mathtt{X}) = \frac{\mathrm{d}}{\mathrm{d}z}W_s(z, \mathtt{X}(z))$ and 
\begin{align}
\frac{\delta \mathcal{W}_s}{\delta \mathtt{X}}[\mathtt{X}'(z)] & =
\int_{\mathbb{R}} \mathrm{d}z\, \frac{\mathrm{d}}{\mathrm{d}z}W_s(\mathtt{X}(z)) 
\nonumber \\ & 
= W_s(\mathtt{x}_{\text{\tiny BP}})- W_s(\Delta_\infty)
\nonumber \\ & 
=\Delta E.
\label{derivativefinalWs}
\end{align}

We now show that the functional derivative of the interaction part in \eqref{decomposition-s-i} does not contribute when $\eta(z,t)$ is replaced 
by $\mathtt{X}'(z)$. By directly applying the definition of the functional derivative, we find
\begin{align}
& \frac{\delta \mathcal{W}_i}{\delta \mathtt{X}}[\mathtt{X}'(z)]
=\int_{\mathbb{R}}\mathrm{d}z\,\Big\{H\big(\mathtt{c}\varoast\big[\lambda^\varoast(\rho^\boxast(\mathtt{X}(z)))
\varoast(\rho^{\prime\boxast}(\mathtt{X}(z))\boxast \mathtt{X}'(z)) \big] \big)
\nonumber \\ &
\,\,\,\,\,\,\,\,-H\big(\mathtt{c}\varoast\big[\lambda^\varoast(\int_0^1\mathrm{d}u\,\rho^\boxast(\mathtt{X}(z+u)))
\varoast(\int_0^1\mathrm{d}s\,\rho^{\prime\boxast}(\mathtt{X}(z+s))\boxast \mathtt{X}'(z+s)) 
\big] \big) \Big\}.
\label{derivative-Wiii}
\end{align}
We notice that the integrand is a total derivative; namely, it is equal to 
\begin{align*}
\frac{1}{L'(1)}\frac{\mathrm{d}}{\mathrm{d}z}\Big\{ & H\big(\mathtt{c}\varoast L^\varoast(\rho^\boxast(\mathtt{X}(z))) \big)
-H\big(\mathtt{c}\varoast L^\varoast(\int_0^1\mathrm{d}u\,\rho^\boxast(\mathtt{X}(z+u)))  \big) \Big\}
\end{align*}
Due to the boundary conditions, we have  
$\lim_{z\to -\infty}\mathtt{X}(z)=  \lim_{z\to -\infty}\mathtt{X}(z+u) = \Delta_\infty$ 
and $\lim_{z\to +\infty}\mathtt{X}(z)=\lim_{z\to +\infty}\mathtt{X}(z+u)= \mathtt{x}_{\text{\tiny BP}}$, and we can conclude that the total derivative 
integrates to zero, thus
\begin{align}\label{derivaWi}
 \frac{\delta \mathcal{W}_i}{\delta \mathtt{X}}[\mathtt{X}'(z)] = 0.
\end{align}

Finally, replacing \eqref{velo}, \eqref{derivativefinalWs}, and \eqref{derivaWi} in \eqref{decomposition-s-i} we get the simple relationship
\begin{align}
 v\int_{\mathbb{R}}\mathrm{d}z\,H\big(\rho^{\prime\boxast}(\mathtt{X}(z))\boxast \mathtt{X}'(z)^{\boxast 2} \big) = \Delta E
\end{align}
which yields the velocity formula \eqref{eqn:vFormulaBMS}.

\section{Applications to specific channels and comparisons with Numerical Experiments}\label{section:velocityCodingApplications}


\subsection{Binary Erasure Channel (BEC)}\label{ssection:vBEC}

The formula for the velocity, when transmission takes place 
over the BEC, can 
be obtained by directly simplifying the general formula in \eqref{eqn:vFormulaBMS}. We note that, since the BEC yields a scalar system, one can also
use the formula for general scalar systems in Section~\ref{section:velocityScalar} (that covers cases beyond coding theory also). We will suppose that 
the underlying code LDPC$(\lambda, \rho)$ is such that the DE equation has a single non-trivial fixed point $x_{\text{\tiny BP}}\neq 0$. Furthermore, we fix 
$\epsilon_{\text{\tiny BP}}\leq \epsilon\leq \epsilon_{\text{\tiny MAP}}$ (recall the channel entropy reduces to $H(\mathtt{c}) = \mathtt{h} =\epsilon$ here).

The channel distribution can be written as $\mathtt{c} = \epsilon\Delta_0 + (1-\epsilon)\Delta_{\infty}$, and the 
profile is of the form $\mathtt{x}(z, t)= x(z, t)\Delta_0 + (1- x(z,t))\Delta_\infty$ where $0\leq x(z,t)\leq 1$ is the scalar erasure probability
at position $z$ and time $t$. This tends to a fixed shape
\begin{align}\label{becshape}
\mathtt{X}(z)=X(z)\Delta_0 + (1-X(z))\Delta_{\infty}.
\end{align}
where $0\leq X(z)\leq 1$ satisfies $\lim_{z\to -\infty} X(z)=0$, $\lim_{z\to +\infty}X(z) = x_{\text{\tiny BP}}$. We have also
\begin{align}\label{derivabecshape}
 \mathtt{X}^\prime(z)=X'(z)\Delta_0 - X'(z)\Delta_{\infty}.
\end{align}
We also note the following identities valid for scalar maps $f, g:\mathbb{R}\to [0,1]$ (such as $\lambda$, $\rho$, $L$, $R$ and their derivatives)
\begin{align}\label{corresp}
\begin{cases}
f^{\varoast}(\mathtt{X}(z)) = f(X(z))\Delta_0 + (1-f(X(z)))\Delta_{\infty},\\
g^{\boxast}(\mathtt{X}(z)) = (1-g(1-X(z))\Delta_0 + g(1-X(z))\Delta_\infty.
\end{cases}
\end{align}

Let us compute the denominator of \eqref{eqn:vFormulaBMS}.
Using \eqref{identityannihilation}, \eqref{derivabecshape}, and 
\eqref{corresp} we have
\begin{align*}
\rho^{\prime\boxast}(\mathtt{X}(z))\boxast\mathtt{X}'(z)^{\boxast 2} &= 
\{(1-\rho'(1-X(z)))\Delta_0+\rho'(1-X(z))\Delta_\infty\}
\boxast\{X'(z)\Delta_0-X'(z)\Delta_\infty \}^{\boxast 2}\\
&=\{(1-\rho'(1-X(z)))\Delta_0+\rho'(1-X(z))\Delta_\infty\}
\boxast\{X'(z)^2\Delta_\infty-X'(z)^2\Delta_0\}\\
&=(1-\rho'(1-X(z)))X'(z)^2\Delta_0
- (1-\rho'(1-X(z)))X'(z)^2\Delta_0\\
&\qquad \qquad \qquad
+\rho'(1-X(z))X'(z)^2\Delta_\infty
-\rho'(1-X(z))X'(z)^2\Delta_0\}\\
&
= \rho'(1-X(z))X'(z)^2\Delta_\infty
- \rho'(1-X(z))X'(z)^2\Delta_0,
\end{align*}
and since $H(\Delta_0)=1$, $H(\Delta_{\infty}) =0$,
and the entropy functional is linear, we obtain the denominator of \eqref{eqn:vFormulaBMS} as
\begin{align*}
H\big(\rho^{\prime\boxast}(\mathtt{X}(z))\boxast\mathtt{X}'(z)^{\boxast 2}\big)
= - \rho^\prime(1-X(z))X'(z)^2 .
\end{align*}
For the numerator of \eqref{eqn:vFormulaBMS}, we have 
$\Delta E = W_{\text{\tiny BEC}}(x_{\text{\tiny BP}})-W_{\text{\tiny BEC}}(0)$, where
the single system potential on the BEC is obtained from \eqref{eqn:singlePotBMS} 
using again \eqref{identityannihilation} and \eqref{corresp}. The exercise yields  
\begin{align*}
W_{\text{\tiny BEC}}(x) & =\frac{1}{R'(1)}(1-R(1-x))-x\rho(1-x)
-\frac{\epsilon}{L'(1)}L(1-\rho(1-x)).
\end{align*}

Putting together these results, the velocity \eqref{eqn:vFormulaBMS}  becomes
\begin{align}
v_{\text{\tiny BEC}}= - \frac{W_{\text{\tiny BEC}}(x_{\text{\tiny BP}})-W_{\text{\tiny BEC}}(0) }
{\int_{\mathbb{R}}\mathrm{d}z\,\rho^\prime(1-X(z))X^\prime(z)^2}.
\label{eqn:formulaBEC}
\end{align}
(Note that, with our normalizations, 
$W_{\text{\tiny BEC}}(0) =0$ for all $\epsilon$.) 
The erasure profile $X(z)$ has to be computed from the one-dimensional integral equation 
\begin{align}
X(z) & - v_{\text{\tiny BEC}}X^\prime(z) 
= \epsilon\int_0^1 \mathrm{d}u\, \lambda\big(1-\int_0^1 \mathrm{d}s\, \rho(1 - X(z-u+s))\big).
\label{eqn:XeqnBEC}
\end{align}
Obviously, the velocity vanishes when $\epsilon\to\epsilon_{\text{\tiny MAP}}$ since then 
$W_{\text{\tiny BEC}}(x_{\text{\tiny BP}})\to W_{\text{\tiny BEC}}(0) =0$.
An important quantity is the slope of the velocity at $\epsilon_{\text{\tiny MAP}}$.
To compute it, we remark that $W_{\text{\tiny BEC}}$ has an explicit dependence on $\epsilon$, as well as an implicit one 
through $x_{\text{\tiny BP}}(\epsilon)$. Thus, 
\begin{align*}
 \frac{\mathrm{d}W_{\text{\tiny BEC}}}{\mathrm{d}\epsilon} & = \frac{\partial W_{\text{\tiny BEC}}}{\partial \epsilon} 
 + \frac{\partial W_{\text{\tiny BEC}}}{\partial x_{\text{\tiny BP}}} \frac{\mathrm{d} x_{\text{\tiny BP}}}{\mathrm{d}\epsilon}
 \nonumber \\ &
 = \frac{\partial W_{\text{\tiny BEC}}}{\partial \epsilon} 
 \nonumber \\ &
 = - \frac{1}{L'(1)}L(1-\rho(1-x_{\text{\tiny BP}})),
\end{align*}
so that, for $\epsilon\to \epsilon_{\text{\tiny MAP}}$, the Taylor expansion to first order yields
\begin{align*}
 W_{\text{\tiny BEC}}(x_{\text{\tiny BP}}) \approx - (\epsilon - \epsilon_{\text{\tiny MAP}}) \frac{1}{L'(1)}L(1-\rho(1-x_{\text{\tiny MAP}})).
\end{align*}
Note that we used $x_{\text{\tiny BP}}(\epsilon) \to x_{\text{\tiny BP}}(\epsilon_{\text{\tiny MAP}}) = x_{\text{\tiny MAP}}$ where $x_{\text{\tiny MAP}}$ is defined as the point 
$x\neq 0$ where the potential is stationary and vanishes.
This yields the linear approximation for the velocity
\begin{align}\label{linear-BEC}
v_l=\frac{(\epsilon-\epsilon_{\text{\tiny MAP}})}{L'(1)}\frac{L(1-\rho(1-x_{\text{\tiny MAP}}))}{\int_{\mathbb{R}}\mathrm{d}z\,\rho'(1-X_{\text{\tiny MAP}}(z))(X_{\text{\tiny MAP}}^\prime(z))^2},
\end{align}
where $X_{\text{\tiny MAP}}(\cdot)$ is the erasure probability 
profile obtained when $\epsilon=\epsilon_{\text{\tiny MAP}}$.

It is interesting to compare \eqref{eqn:formulaBEC} with the upper bound of Theorem 1 in 
\cite{aref2013convergence} for a {\it discrete} system
\begin{align}\label{upperboundBEC}
v_{\text{\tiny B}}=\alpha\frac{W_{\text{\tiny BEC}}(x_{\text{\tiny BP}}) 
- W_{\text{\tiny BEC}}(0)}{\sum\limits_{z\in\mathbb{Z}}\rho^\prime(1-x_z)(x_z-x_{z-1})^2}, \qquad \alpha \leq 2.
\end{align}
In \cite{aref2013convergence} the 
derivation of the bound yields $\alpha\leq 2$ (for $\len$ and $w$ large enough) 
but it is conjectured based on numerical simulations 
that $\alpha=1$ would be a tight bound. Obviously, \eqref{eqn:formulaBEC} and \eqref{upperboundBEC} are consistent. 
We note for reference that another upper bound is also derived in \cite{aref2013convergence}, namely
\begin{align*}
v_{B_2}=\frac{\alpha (W_{\text{\tiny BEC}}(x_{\text{\tiny BP}};\epsilon) 
- W_{\text{\tiny BEC}}(0;\epsilon))}{2W_{\text{\tiny BP}}(x_u;\epsilon)-W_{\text{\tiny BP}}(x_{\text{\tiny BP}};\epsilon)},
\end{align*}
where and $x_u$ and $x_{\text{\tiny BP}}$ are respectively 
the non-trivial unstable and stable fixed points of the potential of 
the uncoupled system $W_s(\cdot;\cdot)$. We do not discuss this further because in practice this turns out to be 
a very loose bound. 

We now compare the {\it analytical} velocity formula \eqref{eqn:formulaBEC} with the {\it empirical}
velocity (called $v_e$ below) obtained by 
simulating the discrete DE equation; we show that it provides 
a very good approximation for the (real) value of the velocity even for 
relatively small values of $w$.  
For the simulations, we consider 
the spatially coupled $(3,6)$ and $(4,6)$-regular 
code ensembles, as well as two irregular LDPC codes (described later). 
We run the simulations for several values of 
the chain length $\len=256,1024$ and the window size $w=3,5,8,16$.
The empirical velocity is the velocity calculated from erasure probability profiles 
of the discrete DE equation. Consider two (discrete) 
profiles $\underline{x}^{(t_1)}$ and $\underline{x}^{(t_2)}$ at any two 
iterations $t_1$ and $t_2$, respectively, with $t_1<t_2$. After the transient phase is over,
the profiles are identical up to translation. 
We call a ``kink" the part of the profile where there is a fast increase 
from $0$ to $x_{\text{\tiny BP}}$ in the erasure probability. The kink ``position'' is the coordinate 
such that the height is equal to $x_{\text{\tiny BP}}/2$, and $\Delta z$ is the difference of two 
such positions (on two different profiles). 
Then, the empirical velocity $v_e$ is 
\begin{align}
v_e=\frac{\Delta z}{w(t_2-t_1)}
\label{eqn:vEmp}
\end{align}
In practice, we get reliable results by taking pairs of profiles separated by $20$ iterations 
and averaging this ratio over every consecutive pair of profiles.
Note that we normalize 
the velocity by $w$ to be able to compare systems with different window widths.

In Table~\ref{table:VdiffW}, we give empirical values $v_e$ of 
the normalized velocities for the spatially 
coupled $(4,6)$-regular code ensemble, with transmission over the BEC(0.6), when 
the spatial length is $1024$ positions and the channel parameter is fixed to $\epsilon=0.6$ (between 
the BP and MAP thresholds), for different values of the window size $w$. We observe 
that the result of our formula $v_{\text{\tiny BEC}}$ gives a 
good estimate of the empirical velocity $v_e$ for all the demonstrated 
values of the window size. We also observe that the linear 
approximation gives a good estimate when the  
 channel parameter is not too far from the MAP threshold $\epsilon_{\text{\tiny MAP}}$. The 
upper bound $v_B$ \cite{aref2013convergence} gives a better estimate as the window size grows larger.

\begin{table}[h]
\caption{Normalized velocities for LDPC($x^4,x^6$) on 
the BEC with a spatial length $1024$, for several $w$ sizes, and $\epsilon=0.6$. The values 
in the table can be compared to 
$v_{\text{\tiny BEC}}=0.0333$ and $v_l=0.0293$.}
\label{table:VdiffW}
\begin{center}
\begin{tabular}{|c||c|c|c|c|}
\hline
  & $w=3$ & $w=5$ & $w=8$ & $w=16$ \\
\hline
$v_e$ & 0.0325 & 0.0335 & 0.0337 & 0.0339  \\
\hline
$v_{B}/\alpha$ & 0.0473 & 0.0410 & 0.0380 & 0.0356 \\
\hline
\end{tabular}
\end{center}
\end{table}

In Table~\ref{table:VdiffE}, we give empirical values of the normalized 
velocities for the spatially coupled $(3,6)$-regular code ensemble, 
with transmission over the BEC($\epsilon$), when the spatial 
length equals $1024$ and the window size $w=8$, for different values 
of the channel parameter $\epsilon$. 
One can compare these values with those in \cite{aref2013convergence} (up 
to a factor equal to $w$ due to the normalization). The result of the 
formula $v_{\text{\tiny BEC}}$ gives the closest estimate to 
the empirical velocity $v_e$ for all values of $\epsilon$.

\begin{table}[h]
\caption{Normalized velocities for
LDPC($x^3,x^6$) on the BEC for spatial length $1024$, $w=8$, and 
several $\epsilon$ values.}
\label{table:VdiffE}
\begin{center}
\begin{tabular}{|c||c|c|c|c|}
\hline
 & $\epsilon=0.45$ & $\epsilon=0.46$ & $\epsilon=0.47$ & $\epsilon=0.48$\\
\hline
$v_e$ & 0.0667 & 0.0458 & 0.0267 & 0.0117 \\
\hline
$v_{\text{\tiny BEC}}$ & 0.0660 & 0.0449 & 0.0272 & 0.0115 \\
\hline
$v_l$ & 0.0506 & 0.0373 & 0.0240 & 0.0108 \\
\hline
$v_{B}/\alpha$ & 0.0781 & 0.0541 & 0.0332 & 0.0142 \\
\hline
$v_{B_2}/\alpha$ & 0.6970 & 0.5008 & 0.3068 & 0.1291 \\
\hline
\end{tabular}
\end{center}
\end{table}

\begin{figure}
\centering
\includegraphics[draft=false,scale=0.27]{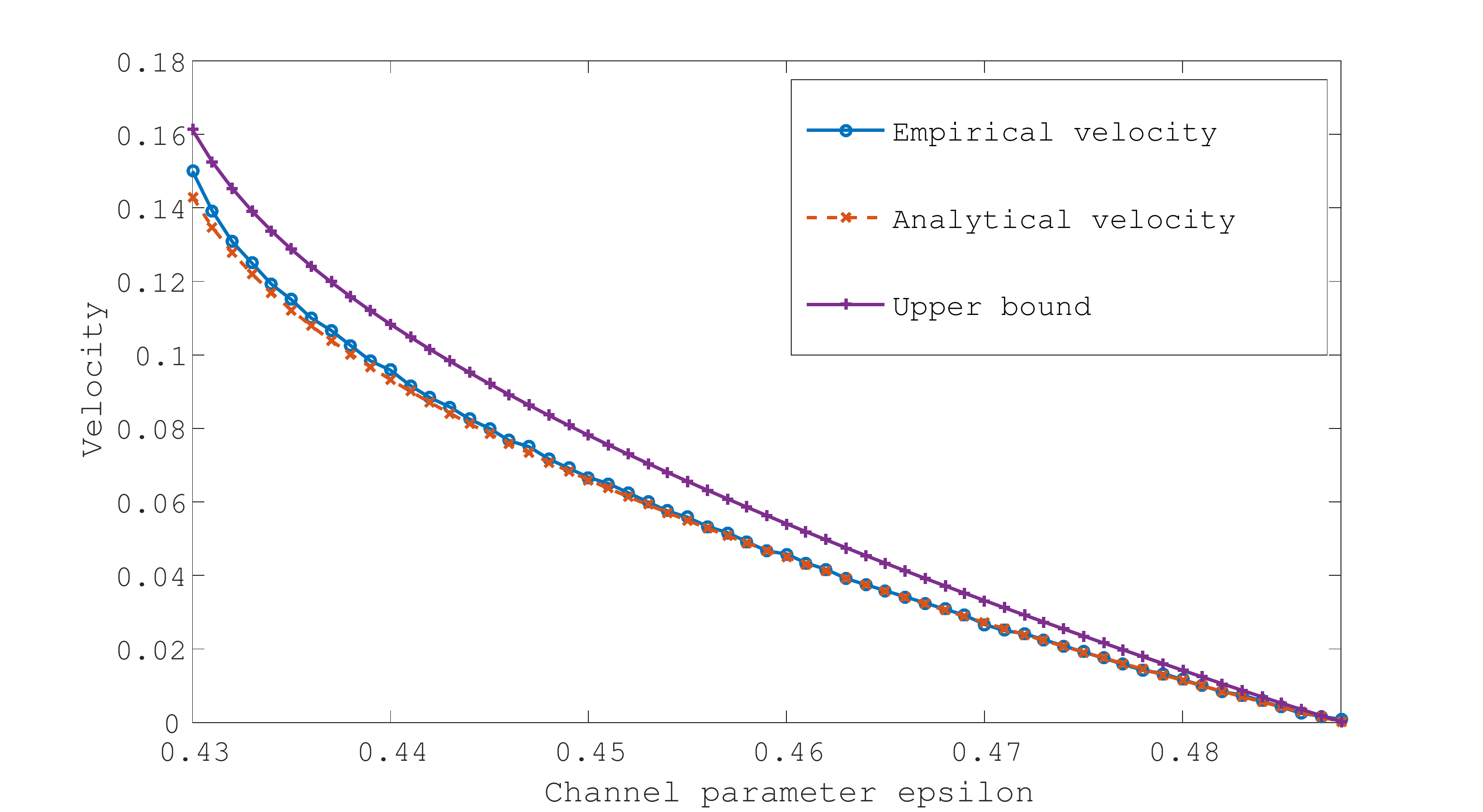}
\caption{Normalized velocities $v_e$, $v_{\text{BEC}}$, 
and $v_{\text{B}}/\alpha$ (in the order of the legend) for 
the $(3,6)$-regular ensemble with spatial length $256$, window size $w=3$, and 
$\epsilon_{\text{\tiny BP}}=0.43<\epsilon<\epsilon_{\text{\tiny MAP}}=0.4881$.}
\label{fig:velocities36}
\end{figure}

\begin{figure}
\centering
\includegraphics[draft=false,scale=0.27]{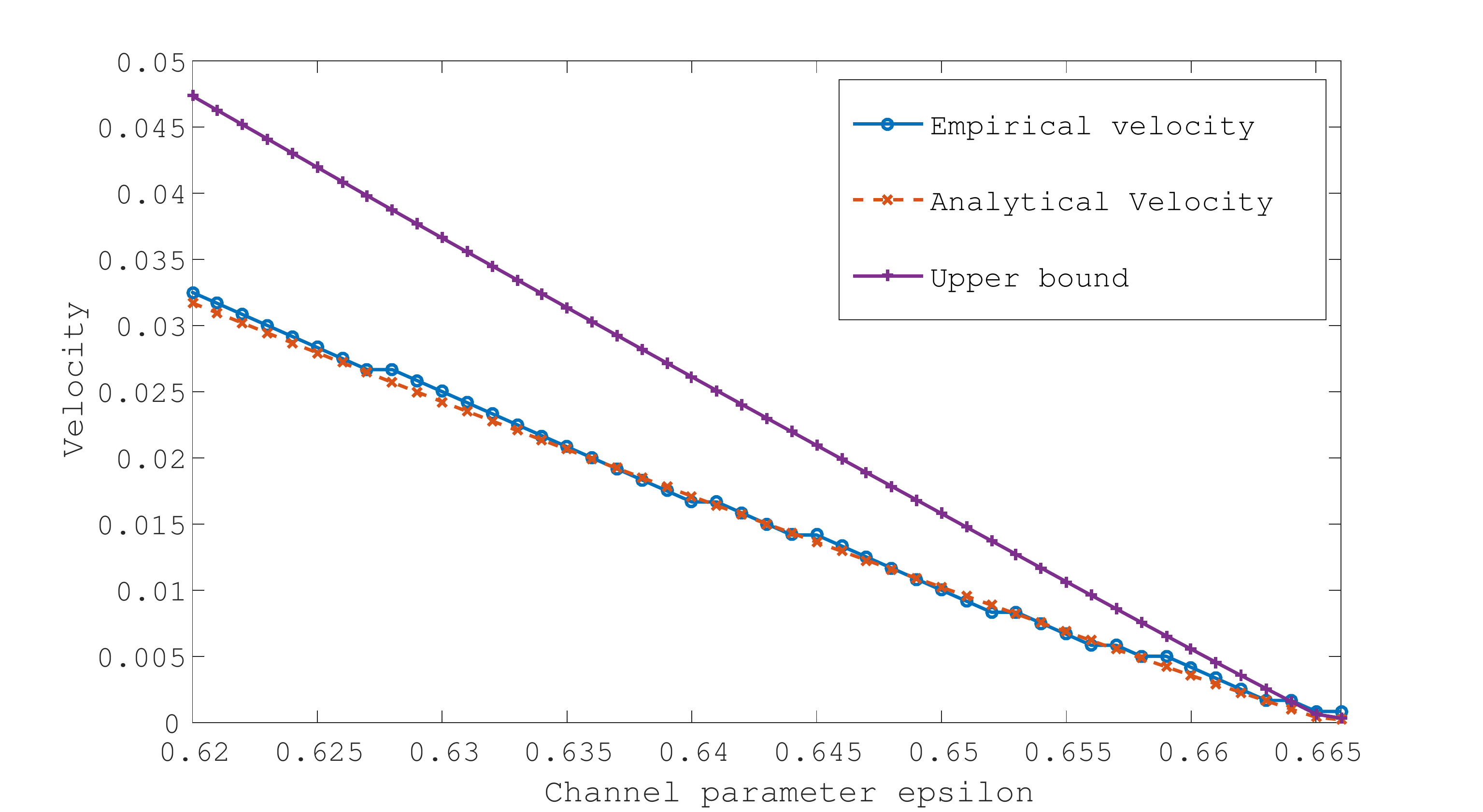}
\caption{Normalized velocities $v_e$, $v_{\text{BEC}}$, and $v_{\text{B}}/\alpha$ 
(in the order of the legend) of the decoding profile for the $(4,6)$-regular ensemble of spatial length
$256$, window size $w=3$, and 
$\epsilon_{\text{\tiny BP}}=0.515<\epsilon<\epsilon_{\text{\tiny MAP}}=0.666$.}
\label{fig:velocities46}
\end{figure}

Figures~\ref{fig:velocities36} and~\ref{fig:velocities46} show 
the empirical velocity $v_e$, the analytical 
velocity $v_{\text{\tiny BEC}}$, and the upper bound $v_B$
for the spatially coupled $(3,6)$-regular 
code ensemble, with spatial length $256$ and window size $w=3$.
We remark 
that our formula fits very well, for the $(3,6)$-regular code, with the empirical velocity for all values of the channel 
parameter $\epsilon\in[\epsilon_{\text{\tiny BP}},\epsilon_{\text{\tiny MAP}}] = [0.43,0.488]$.
The agreement is quite good also for the $(4,6)$-regular code and very good for more than half of the interval 
$[\epsilon_{\text{\tiny BP}},\epsilon_{\text{\tiny MAP}}] \approx [0.515,0.719]$.

\begin{figure}
\centering
\includegraphics[draft=false,scale=0.27]{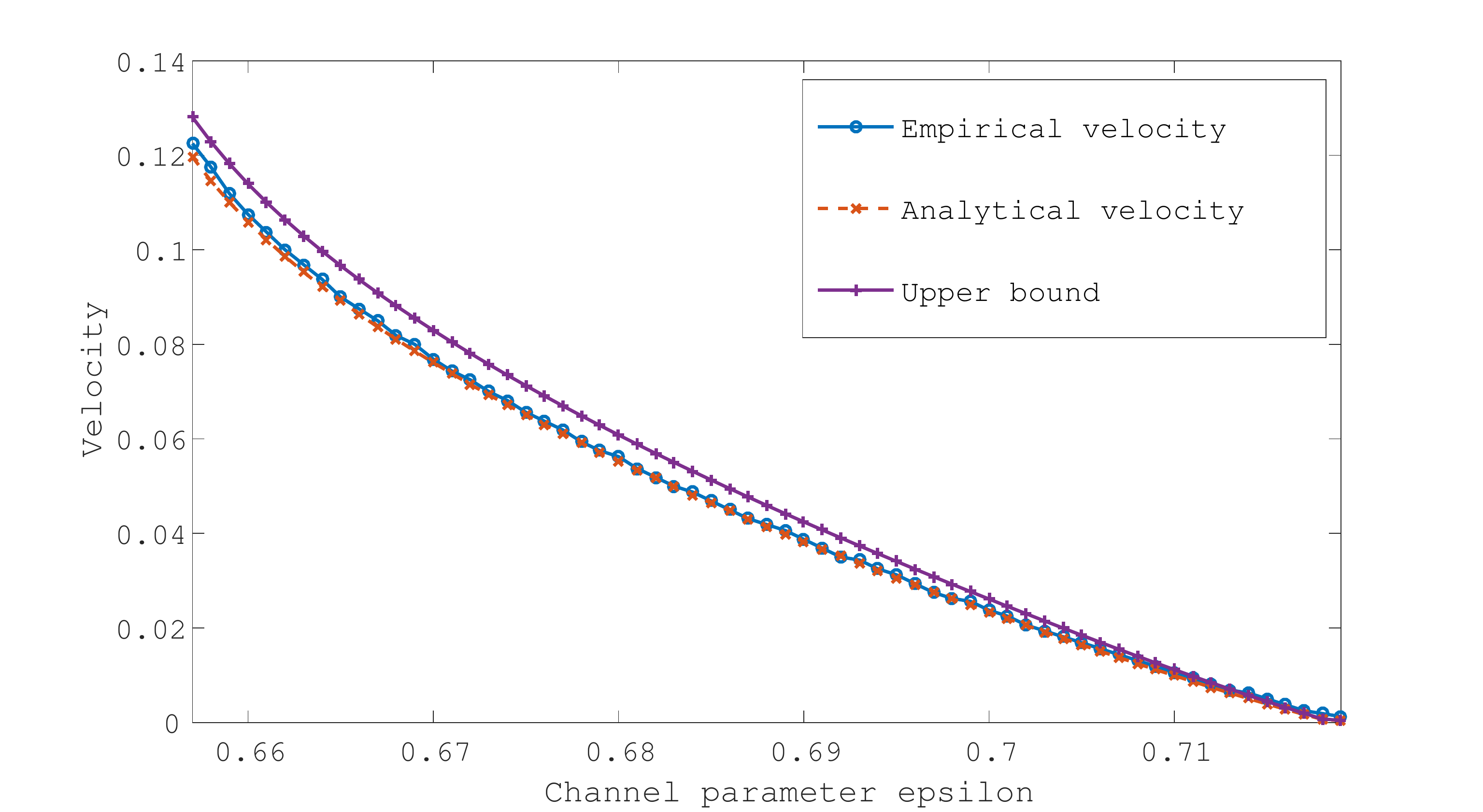}
\caption{Normalized velocities $v_{\text{\tiny BEC}}$, $v_e$, and $v_B/\alpha$ for an ensemble with 
$L(x)=0.3x^2+0.6x^3+0.1x^5$, $R(x)=x^4$, spatial length $1024$, window size $w=4$, and $\epsilon_{\text{\tiny BP}}=0.657<\epsilon<\epsilon_{\text{\tiny MAP}}=0.719$.}
\label{fig:velocitiesIrr1}
\end{figure}

\begin{figure}
\centering
\includegraphics[draft=false,scale=0.27]{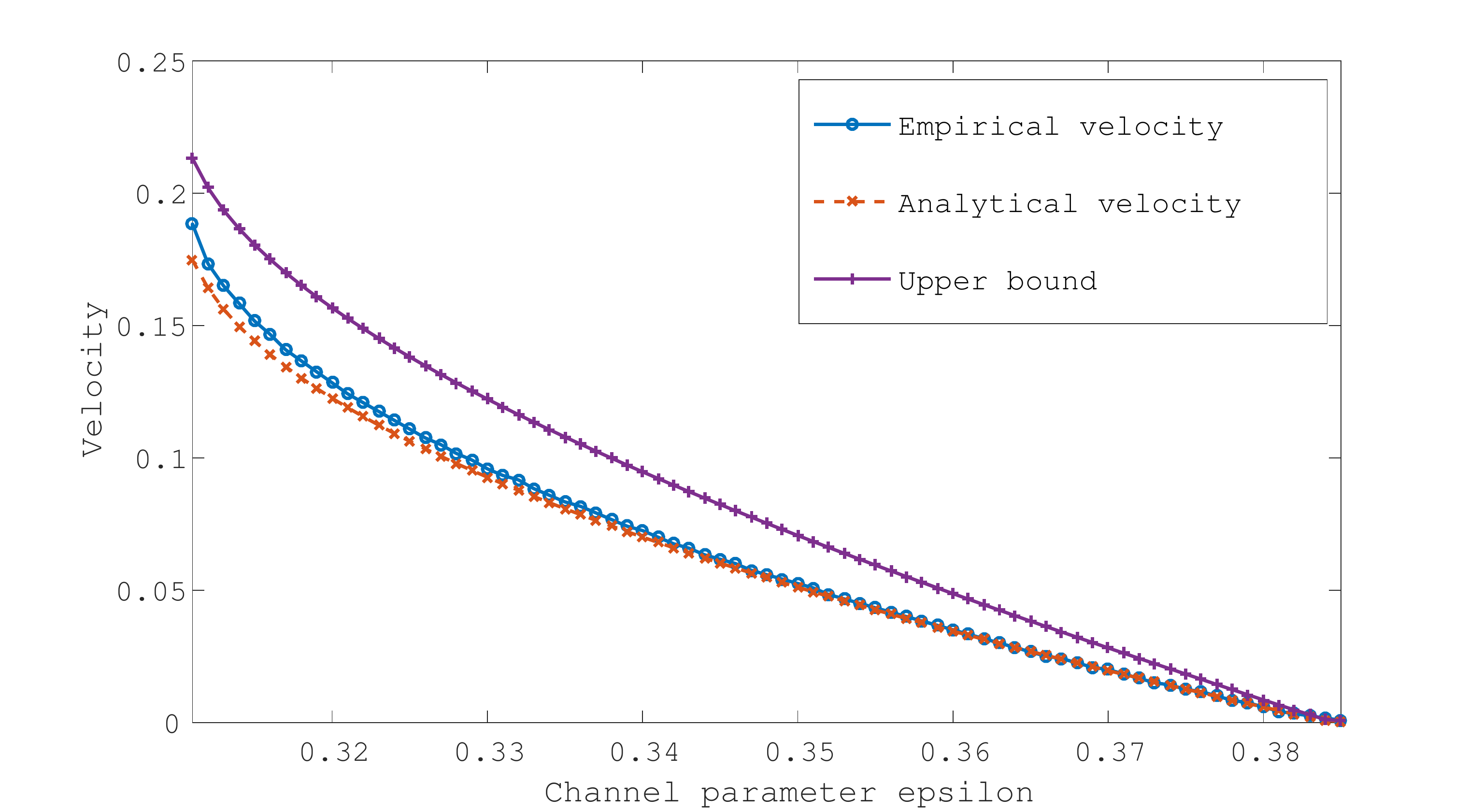}
\caption{Normalized velocities $v_{\text{\tiny BEC}}$, $v_e$, $v_B/\alpha$ for an ensemble with 
$L(x)=0.4x^3+0.3x^4+0.3x^5$, $R(x)=0.5x^8+0.5x^{12}$, spatial length $256$, window size $w=3$, and 
$\epsilon_{\text{\tiny BP}}=0.311<\epsilon<\epsilon_{\text{\tiny MAP}} = 0.385.$}
\label{fig:velocitiesIrr2}
\end{figure}

We also illustrate the results for two irregular code ensembles in  Figures ~\ref{fig:velocitiesIrr1} and 
\ref{fig:velocitiesIrr2}. The first one has node degree distributions $L(x)=0.3x^2+0.6x^3+0.1x^5$ and $R(x)=x^4$, 
spatial length $1024$, and window size $w=4$. The agreement between $v_{\text{\tiny BEC}}$ and $v_e$ is excellent for the whole range
$\epsilon\in[\epsilon_{\text{\tiny BP}},\epsilon_{\text{\tiny MAP}}] = [0.657,0.719]$. 
The second one has  $L(x)=0.4x^3+0.3x^4+0.3x^5$ and $R(x)=0.5x^8+0.5x^{12}$, spatial length $256$, and window 
size $w=3$. The agreement between the velocities is also very good for most of the range 
$\epsilon\in[\epsilon_{\text{\tiny BP}},\epsilon_{\text{\tiny MAP}}] = [0.311,0.385]$.

\subsection{Gaussian Approximation (GA)}\label{ssection:vGA}

DE equations relate probability densities and as such we may need to track an infinite set of parameters (except for the BEC where 
the space of densities can be parametrized by a single real number). In many situations, such as the case when we have large degrees, for example, the 
densities are well approximated by 
Gaussians, which enables us to project the DE equations down to a low dimensional space. There are several variants of the Gaussian approximation 
(see for example \cite{chung2001analysis}, \cite{chung2001capacity}, \cite{chung2000construction}), and here we use it in a form called the ``reciprocal channel approximation''
proposed in \cite{chung2001capacity}, \cite{chung2000construction}. 

The idea is to 
assume that the densities of the LLR messages appearing in the DE equations are symmetric Gaussian densities. 
Such densities take the form
\begin{align}
\mathrm{d}\mathtt{x}(\alpha)=\frac{\mathrm{d}\alpha}{\sqrt{2\pi\sigma^2}}\exp\Big(-\frac{(\alpha -m)^2}{2\sigma^2}\Big),
\end{align}
with the mean $m$ and variance $\sigma^2$ satisfying $\sigma^2 = 2m$.  
Furthermore, the channel density $\mathtt{c}$ is replaced by that corresponding to a BIAWGNC($\sigma_n^2$)
with the same entropy $H(\mathtt{c})$. 
Density evolution can then conveniently be expressed
in terms of the entropies $p_z^{(t)}=H(\mathtt{x}_z^{(t)})$. This is done as follows.
Let $\psi(m)$ denote the entropy\footnote{For indications on the numerical implementation 
of this function see \cite{richardson2008modern}, pp.194 and 237.}  of a symmetric Gaussian density
of mean $m$ given by 
\begin{align}
\psi(m)=\frac{1}{\sqrt{4\pi m}}\int_{\mathbb{R}}\mathrm{d}z\,e^{-\frac{(z-m)^2}{4m}}\log_2(1+e^{-z}).
\end{align}
Thus $\psi^{-1}(p)$ denotes the mean of a symmetric Gaussian density $\mathtt{x}$ of entropy $p = H(\mathtt{x})$. Take two 
symmetric Gaussian densities $\mathtt{x}_1$ and $\mathtt{x}_2$ of means $m_1$ and $m_2$ and entropies $p_1 =
\psi(m_1)$ and $p_2 =\psi(m_2)$. We have, 
in general, 
\begin{align}\label{firstGA}
H(\mathtt{x}_1\varoast\mathtt{x}_2) = \psi(\psi^{-1}(p_1) + \psi^{-1}(p_2)),
\end{align}
which just expresses the fact that a usual convolution of two Gaussian densities of means $m_1$ and $m_2$ is a Gaussian density 
of mean $m_1+m_2$. On the other hand $\mathtt{x}_1\boxast \mathtt{x}_2$ is not exactly Gaussian so there is no exact formula but the idea here is 
to preserve the duality rule
$H(\mathtt{x}_1\boxast \mathtt{x}_2) + H(\mathtt{x}_1\otimes \mathtt{x}_2) = H(\mathtt{x}_1) + H(\mathtt{x}_2)$. Writing this relation as 
$$
1- H(\mathtt{x}_1\boxast \mathtt{x}_2) = H((\Delta_0-\mathtt{x}_1)\varoast (\Delta_0- \mathtt{x}_2)),
$$ 
and noting that $H(\Delta_0-\mathtt{x}_1)=1-p_1$,
$H(\Delta_0-\mathtt{x}_2)=1-p_2$ suggests the approximation 
\begin{align}\label{secondGA}
H(\mathtt{x}_1\boxast \mathtt{x}_2) = 1- \psi(\psi^{-1}(1-p_1) + \psi^{-1}(1-p_2)).
\end{align}

Looking at the entropies of the DE equations \eqref{firstGA} and \eqref{secondGA} imply
(we will limit ourselves to regular codes for simplicity)
\begin{align}
\begin{cases}
H(\mathtt{x}^{(t)})=H(\mathtt{c}\varoast\mathtt{y}^{{(t)}\varoast(\ell-1)}),\\
H(\mathtt{y}^{(t+1)})=H(\mathtt{x}^{(t)\boxast(r-1)}).
\end{cases}
\end{align}
and setting $p^{(t)} = H(\mathtt{x}^{(t)})$, $q^{(t)}=H(\mathtt{y}^{(t)}$ we find
\begin{align}\label{eqn:DEentropies}
\begin{cases}
p^{(t)}=\psi\big(\psi^{-1}(H(\mathtt{c}))+(\ell-1)\psi^{-1}(q^{(t)})\big),\\
q^{(t+1)}=1-\psi\big((r-1)\psi^{-1}(1-p^{(t)})\big).
\end{cases}
\end{align}
These equations can be combined into 
\begin{align}
p^{(t+1)}=\psi\Big((\ell-1)\psi^{-1}\big( 1-\psi&((r-1)\psi^{-1}(1-p^{(t)})) \big)
+\psi^{-1}(H(\mathtt{c}))\label{eqn:DEentropy} \Big).
\end{align}
The corresponding potential function is easily found from \eqref{eqn:singlePotBMS}
\begin{align}
W_{\text{\tiny GA}}(p) = \frac{1}{r}\big(1- \psi\big(&r\psi^{-1}(1-p)\big)\big)
+\psi\big(r\psi^{-1}(1-p)\big) -\psi\big((r-1)\psi^{-1}(1-p)\big)
\nonumber \\ &
-\frac{1}{\ell}\psi\big(\psi^{-1}(H(\mathtt{c}))+\ell\psi^{-1}\big(1-\psi((r-1)\psi^{-1}(1-p)) \big) \big).
\end{align}

\begin{figure}
\centering
\includegraphics[draft=false,scale=0.3]{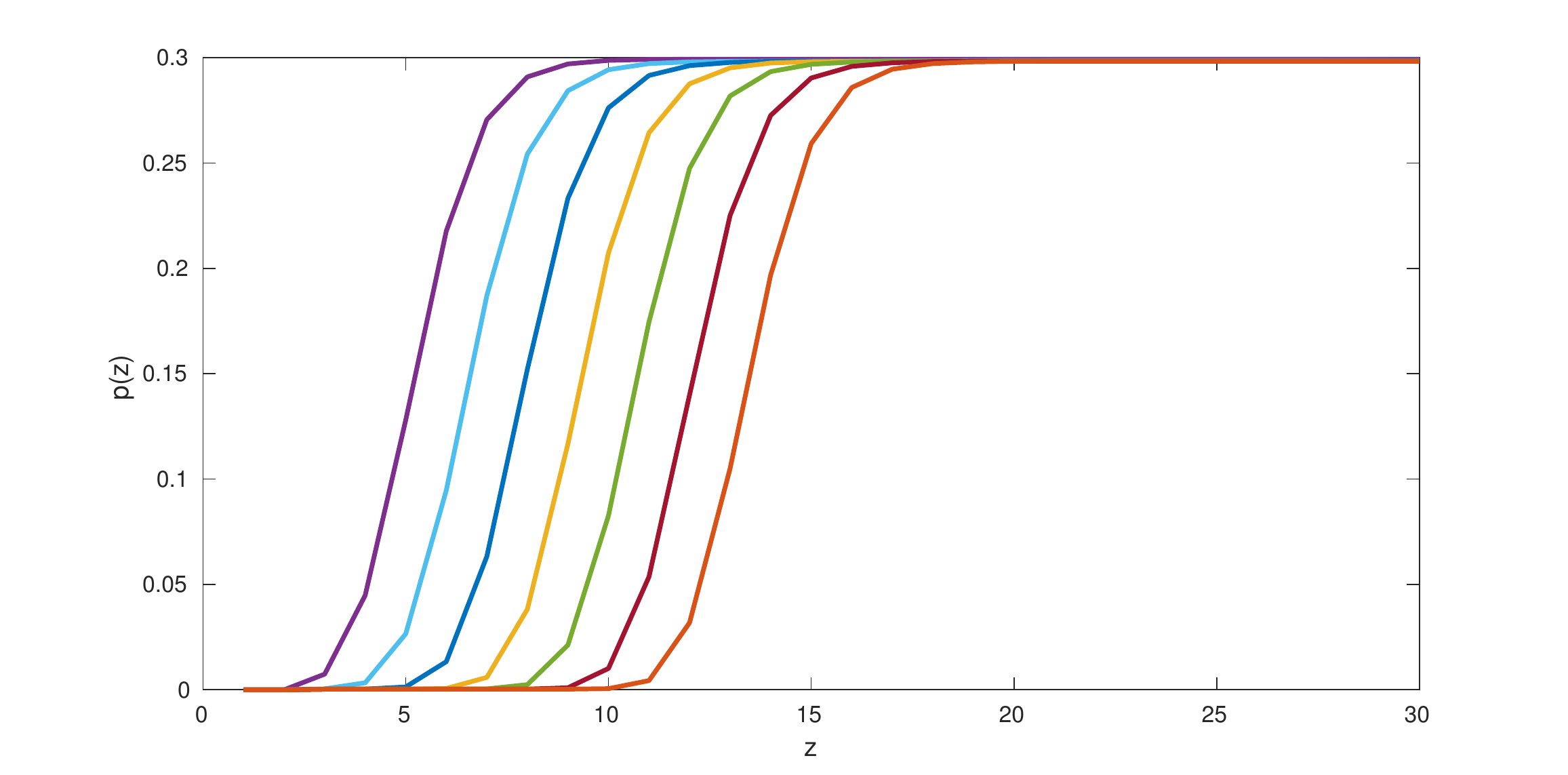}
\caption{The profile of 
entropies $p(z,t)$ plotted every 10 iterations starting from iteration 20. 
We take the $(3,6)$-regular LDPC code with chain length $\len=30$, window size $w=3$, and we consider the BIAWGN channel 
with mean $\psi^{-1}(H(\mathtt{c}))=2.4$.}
\label{fig:waveGA}
\end{figure}

For the coupled system,
we denote by $p_z$ the {\it average over positions $\{z, \dots, z+w\}$} of the entropy of  
symmetric Gaussian densities emanating from the variable nodes. The coupled DE equations then take the form
\begin{align}
p_z^{(t+1)}& =\frac{1}{w}\sum\limits_{i=0}^{w-1}\psi\Big((\ell-1)\psi^{-1}\Big(\frac{1}{w}\sum\limits_{j=0}^{w-1}\big( 1 - \psi((r-1)\psi^{-1}(1-p_{z-i+j}^{(t)}))\big) \Big) +\psi^{-1}(H(\mathtt{c}))\Big).
\end{align}
This coupled recursion can be solved with appropriate boundary conditions and one  
observes a scalar wave propagation, as shown in Figure~\ref{fig:waveGA}. 

We are now ready to discuss the application to the velocity formula. 
The continuum limit 
is obtained exactly as in Section~\ref{ssection:continuumLimit}. The assumption that the density $\mathtt{x}(z, t)$ tends to a fixed shape
$\mathtt{X}(z-v_{\text{\tiny GA}}t)$ after the transient phase implies that its entropy $p(z ,t)$ tends to 
$P(z - v_{\text{\tiny GA}}t) \equiv H(\mathtt{X}(z-v_{\text{\tiny GA}}t))$, where $P(z)$ is a scalar function (independent of initial conditions)
satisfying the integral equation
\begin{align}
&P(z) - v_{\text{\tiny GA}}P^\prime(z) = 
\int_0^1\mathrm{d}u\, \psi\Big(\psi^{-1}(H(\mathtt{c})) +(\ell-1)
(\ell-1) \psi^{-1}\big(1-\int_0^1 \mathrm{d}s\, \psi\big((r-1)\psi^{-1}(1-P(z-u+s))\big)\big)\Big).
\end{align}
and the boundary conditions $\lim_{z\to -\infty}P(z)=0$ and $\lim_{z\to +\infty}=p_{\text{\tiny BP}}$ where $p_{\text{\tiny BP}}$ is the non-trivial 
fixed point of \eqref{eqn:DEentropy}. We will now show that the velocity formula reduces to 
\begin{align}
v_{\text{\tiny GA}}=-\frac{W_{\text{\tiny GA}}(p_{\text{\tiny BP}})-W_{\text{ \tiny GA}}(0)}{(r-1)\int_\mathbb{R}\mathrm{d}z\,
P^\prime(z)^2\frac{\psi''((r-2)\psi^{-1}(1-P(z)))}{\psi'(\psi^{-1}(1-P(z)))^2}}.
\label{eqn:velocityGauss}
\end{align}

To derive \eqref{eqn:velocityGauss}, we consider the denominator in \eqref{eqn:vFormulaBMS}
and write it as follows
\begin{align}
(r-1)H  \big(\mathtt{x}(z)^{\boxast(r-2)}\boxast&\mathtt{x}'(z)^{\boxast 2} \big)
= (r-1)\lim_{\delta\to 0}\frac{1}{\delta^2}\Big\{H\big(\mathtt{x}(z)^{\boxast(r-2)}\boxast\mathtt{x}(z+\delta)^{\boxast 2} \big) 
\nonumber \\& 
-2H\big(\mathtt{x}(z)^{\boxast(r-2)}\boxast\mathtt{x}(z+\delta)\boxast\mathtt{x}(z) \big)
+H\big(\mathtt{x}(z)^{\boxast(r-2)}\boxast\mathtt{x}(z)^{\boxast 2} \big) \Big\}.
\label{entropies-GA-delta1}
\end{align}
Computing each entropy in the Gaussian approximation, we find for the bracket on the right-hand side
\begin{align}
\frac{1}{\delta^2}&\Big\{\big[1-\psi((r-2)\psi^{-1}(1-p(z))+2\psi^{-1}(1-p(z+\delta))) \big]
\nonumber \\
& -2\big[1-\psi((r-1)\psi^{-1}(1-p(z))+\psi^{-1}(1-p(z+\delta))) \big]
 +\big[1-\psi(r\psi^{-1}(1-p(z))) \big]\Big\}.
\label{entropies-GA-delta2}
\end{align}
In Appendix~\ref{appendixTaylor}, we compute the limit of this term when $\delta\to 0$ by appropriate Taylor expansions and find
\begin{align}
&-\Bigg(\frac{\mathrm{d}P(z)}{\mathrm{d}z}\Bigg)^2\frac{\psi''\big((r-2)\psi^{-1}(1-P(z))\big)}{\Big(\psi'(\psi^{-1}(1-P(z)))\Big)^2}.
\label{final-for-ga-velocity}
\end{align}
This concludes the derivation of \eqref{eqn:velocityGauss}.

\begin{figure}
\centering
\includegraphics[draft=false,scale=0.27]{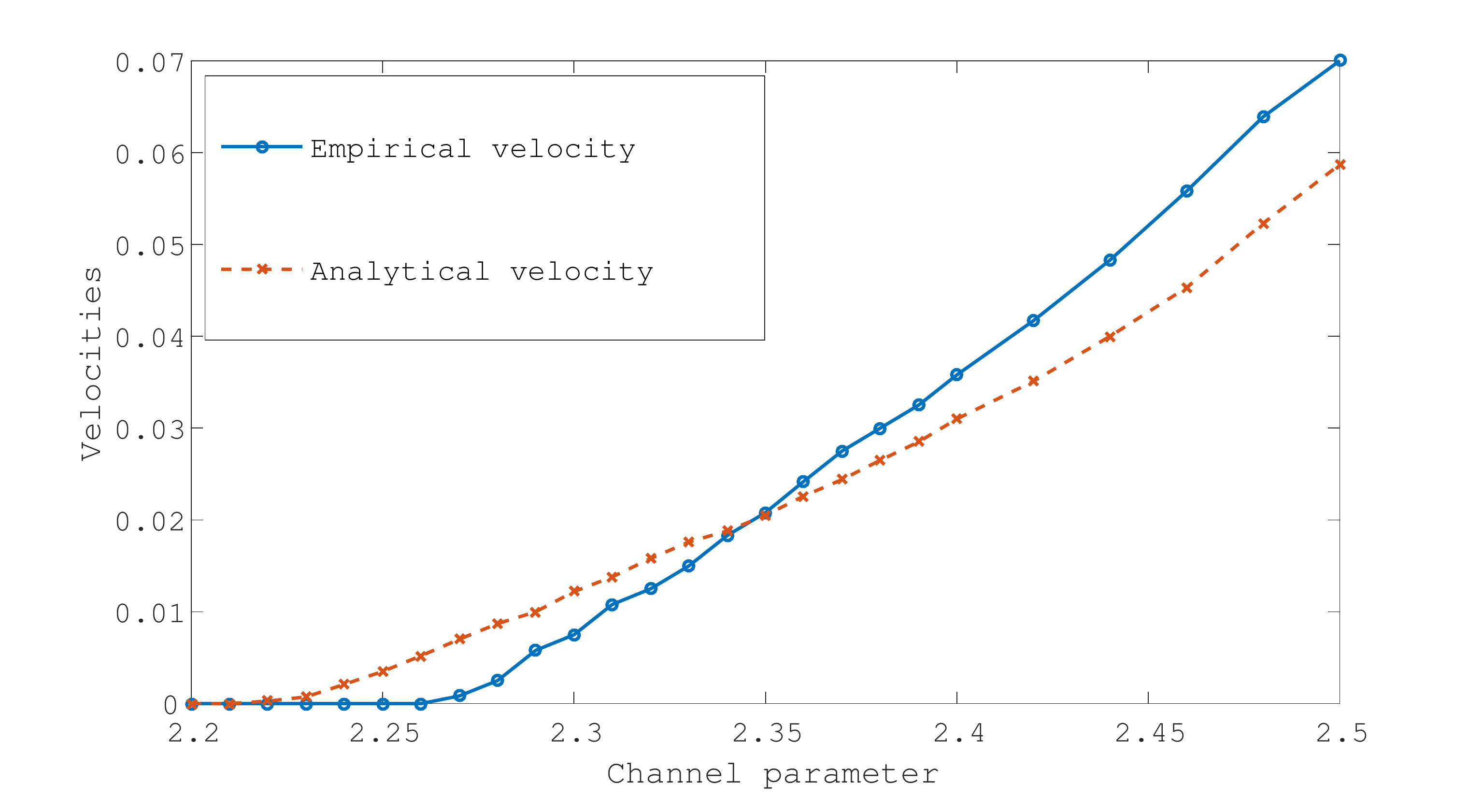}
\caption{Normalized velocities $v_{\text{\tiny GA}}$ and $v_e$ for the spatially coupled $(3,6)$-regular ensemble within the Gaussian approximation framework, for a spatial length of $100$ and $w=3$, as a function of $\psi^{-1}(H(\mathtt{c}))=2/\sigma_n^2$.}
\label{fig:velocitiesGA}
\end{figure}

Table~\ref{table:vGauss} gives a comparison of 
analytical and empirical velocities, $v_{\text{\tiny GA}}$ and 
$v_{e,\text{\tiny GA}}$, obtained for the $(3,6)$ and the $(4,8)$-regular ensembles, for a spatial length of $100$ and a window size $w=3$ for 
different values of $\psi^{-1}(H(c)) = \sigma_n^2/2$ (twice the signal to noise ratio). We also plot both velocities for the $(3,6)$-regular ensemble for the same parameters in Figure~\ref{fig:velocitiesGA}. We conjecture that the errors incurred from these plots are due to numerical errors involved in computing the functions $\psi$ and its inverse.

\begin{table}[htb]
\caption{Normalized velocities of the waves within the Gaussian approximation on the $(3,6)$ and $(4,8)$-regular code ensembles 
 with $\len=100$, $w=3$.
}
\label{table:vGauss}
\begin{center}
\begin{tabular}{|c||c|c|c|c|}
\hline
{\bf $2/\sigma_n^2$} & {\bf 2.33} & {\bf 2.35} & {\bf 2.38} & {\bf 2.40}\\
\hline
$v_{\text{\tiny GA}}$\, $(3,6)$  & 0.0176 & 0.0205 & 0.0265 & 0.0310\\
\hline
$v_e$\, $(3,6)$ & 0.0150 & 0.0208 & 0.0300 & 0.0358 \\
\hline
$v_{\text{\tiny GA}}$\, $(4,8)$ & 0.0237 & 0.0258 & 0.0312 & 0.0381 \\
\hline
$v_e$\, $(4,8)$ & 0.0217  & 0.0250 & 0.0308 & 0.0342  \\
\hline
\end{tabular}
\end{center}
\end{table}

\section{Application to Scaling Laws for Finite-Length Coupled Codes}\label{section:scalingLaw}

The authors in \cite{olmos2014scaling} propose a scaling law to predict the error probability of a finite-length spatially coupled $(\ell,r,\len)$ code when transmission takes place over the BEC. 
The derived scaling law depends on {\it scaling parameters}, one of which we will relate to the velocity of the decoding wave. 
The $(\ell,r,\len)$ ensemble considered in \cite{olmos2014scaling} differs slightly from the purely random ensemble we consider in this work. However, as we will see, our formula for the velocity yields results that are reasonably good for this application. We briefly describe this ensemble and the scaling law.

The $(\ell,r,\len)$ ensemble combines the benefits of purely random codes (that we consider in this work) and protograph-based codes \cite{thorpe2003low}. The randomness involved in the construction makes the ensemble relatively easy to analyze, and the structure added to the construction due to its similarity to protograph-based codes improves the performance of the code.
The ensemble is constructed as follows: Make $\len+w$ copies of an uncoupled code at positions $z=-w+1,\dots,\len$. All edges are erased then reconnected such that a variable node at position $z_0$ has exactly one edge with each set of check nodes at positions $z_0+i$, where $i=0,\dots,\ell-1$. The check nodes are chosen such that the regularity of their degree is maintained. Therefore, every variable node has $\ell$ emanating edges and every check node has $r$ such edges.

We consider transmission over the BEC. In this case, the BP decoder can be seen as a peeling decoder \cite{stinner2016finite}. Whenever a variable node is decoded, it is removed from the graph along with its edges. One way to track this peeling process is to analyze the evolution of the degree distribution of the residual graph across iterations, which serves as a sufficient statistic. This statistic can be described by a system of differential equations, whose solution determines the mean and variance of the {\it fraction of degree-one check nodes} and the variance around this mean at any time during the decoding process. We call $\hat r_1$
the mean.

It has been shown in \cite{olmos2014scaling} that there exists a  {\it steady state phase} where the mean and the variance are constant. It is exactly during this phase that one can observe the progression of the soliton. We note that here we consider one-sided termination instead of two-sided termination (as considered in \cite{olmos2014scaling}), so the fraction $\hat r_1$ here is equal to half the fraction called $\hat r_1(*)$ in \cite{olmos2014scaling}.

Let $\epsilon_{(\ell,r,\len)}$ denote the BP threshold of the finite-size $(\ell,r,\len)$ ensemble (for large $\len$ this is close 
to $\epsilon_{\text{\tiny MAP}}$ due to threshold saturation). We can write the first-order Taylor expansion of $\hat{r}_1\big|_\epsilon$ around $\epsilon < \epsilon_{(\ell,r,\len)}$ as
\begin{align*}
\hat{r}_1\big|_\epsilon\approx\hat{r}_1\big|_{\epsilon_{(\ell,r,\len)}}+\gamma\Delta\epsilon +
\mathcal{O}(\Delta\epsilon^2).
\end{align*}
where $\Delta\epsilon=\epsilon_{(\ell,r,\len)}-\epsilon$.
Thus, for a given $\epsilon<\epsilon_{(\ell,r,\len)}$ and since $\hat{r}_1\big|_{\epsilon_{(\ell,r,\len)}}=0$ (by definition), we obtain
$$\gamma\approx\frac{\hat{r}_1\big|_\epsilon}{\Delta\epsilon}.$$
This parameter $\gamma$ enters in the scaling law and is therefore quite important. So far, it has been determined only  experimentally. 
It would clearly be desirable to have a theoretical handle on $\gamma$. It is argued in \cite{olmos2014scaling} that $\gamma\approx\bar\gamma$ where $\bar\gamma=x_{\text{\tiny BP}}/c$ and $c$ is a real positive constant that behaves like $\Delta\epsilon / v_{\text{\tiny BEC}}$, i.e.,
\begin{align}\label{eqn:gamma2}
\bar\gamma\approx\frac{x_{\text{\tiny BP}}\,v_{\text{\tiny BEC}}}{\Delta\epsilon}
\end{align}
It is expected that this formula becomes exact in an asymptotic limit where threshold saturation takes place
$\epsilon_{(\ell,r,\len)}\to \epsilon_{\text{\tiny MAP}}$. Using  the linearization \eqref{linear-BEC}, we obtain
\begin{align}
\bar\gamma \to 
\frac{x_{\text{\tiny MAP}}}{L'(1)}
\frac{L(1-\rho(1-x_{\text{\tiny MAP}}))}{\int_{\mathbb{R}}\mathrm{d}z\,\rho'(1-X_{\text{\tiny MAP}}(z))(X_{\text{\tiny MAP}}^\prime(z))^2}.
\end{align}
The parameter $\bar\gamma$ is simply equal to the erasure probability times the slope of the velocity 
at $\epsilon_{\text{\tiny MAP}}$. 

We compare  the values of $\gamma$ and $\bar\gamma$ for different values of $\ell$ and $r$, at a channel parameter $\epsilon=\epsilon_{(\ell,r,\len)}-0.04$, in Table~\ref{table:gamma}. The experimental values of $\gamma$ are taken from \cite{olmos2014scaling} and, for those of $\bar\gamma$, we use the analytical velocity \eqref{eqn:formulaBEC}. We observe that the numbers roughly agree. There are two reasons that can explain the discrepancies. Firstly, we derive the velocity for the purely random spatially coupled graph ensemble whereas
the ensemble considered in \cite{olmos2014scaling} is more structured. Note also that as $\ell$, $r$ increase,
the window size of the structured ensemble increases, so the finite size effects at fixed $\len=100$ may be more marked. Secondly, expression
\eqref{eqn:gamma2} is valid when $\epsilon \to \epsilon_{(\ell,r,\len)}$, whereas in~\ref{table:gamma} 
$\Delta\epsilon=0.04$, which is relatively large (this choice in \cite{olmos2014scaling} is due to stability issues 
in numerical integration techniques when $\epsilon \to \epsilon_{(\ell,r,\len)}$). We conjecture that the second issue 
is the dominant reason for the difference between the values of $\gamma$ and $\bar\gamma$ and that, in fact, the velocity for the structured ensemble is not very different from the one predicted by our formula \eqref{eqn:formulaBEC}.

\begin{table}[h]
\caption{Values of $\gamma$ and $\bar{\gamma}$ for $\epsilon_{(\ell,r,L)}-\epsilon=0.04$, $\len=100$, and several values of $\ell$ and $r$}
\label{table:gamma}
\begin{center}
\begin{tabular}{|c|c|c|c|c|}
\hline
l  & r & $\epsilon_{\text{\tiny MAP}}$ & $\gamma$ & $\bar\gamma$ \\
\hline
3 & 6 & 0.4881 & 2.155 & 1.960  \\
\hline
4 & 8 & 0.4977 & 2.120 & 1.779 \\
\hline
5 & 10 & 0.4994 & 2.095 & 1.733 \\
\hline
6 & 12 & 0.4999 & 2.075 & 1.722 \\
\hline
4 & 12 & 0.3302 & 2.140 & 1.778 \\
\hline
5 & 15 & 0.3325 & 2.115 & 1.746 \\
\hline
4 & 6 & 0.6656 & 2.100 & 1.735 \\
\hline
\end{tabular}
\end{center}
\end{table}

\section{Velocity for General Scalar Coupled Systems}\label{section:velocityScalar}

In this section, we consider general scalar spatially coupled systems. That is, we do not restrict ourselves to coding problems; however, we only consider systems in which the ``density evolution type" analysis of message passing algorithms involves \emph{scalar} values. Our main result is again a general formula for the velocity of the soliton in the framework of general scalar coupled systems.
There are numerous systems that fall in this class - coding with transmission on the BEC being one of the simplest - and in Sections~\ref{ssection:GLDPC},~\ref{ssection:CS} we will illustrate our results with two examples, namely
generalized LDPC codes on the BEC and BSC, as well as compressive sensing. 

\subsection{General scalar systems}\label{ssection:vScalarNotation}

We adopt the framework and notation in \cite{yedla2014simple}.
We denote by $\epsilon\in [0,\epsilon_{\text{\tiny max}}]$ where $\epsilon_{\text{\tiny max}}\in(0,\infty)$ the interval of values for the control parameter $\epsilon$.
Consider bounded, smooth functions that are increasing in both their arguments $g:[0,x_{\text{\tiny max}}(\epsilon)]\times[0,\epsilon_{\text{\tiny max}}]\to[0,y_{\text{\tiny max}}(\epsilon)]$ and $f:[0,y_{\text{\tiny max}}(\epsilon)]\times[0,\epsilon_{\text{\tiny max}}]\to [0,x_{\text{\tiny max}}(\epsilon)]$ where 
$x_{\text{\tiny max}}(\epsilon)$, $y_{\text{\tiny max}}(\epsilon)\in(0,\infty)$ and $y_{\text{\tiny max}}(\epsilon)=g(x_{\text{\tiny max}}(\epsilon);\epsilon)$.
The scalar recursions that interest us are
\begin{align}
x^{(t+1)}= f( g(x^{(t)};\epsilon);\epsilon ),
\label{eqn:scalarDEuncoupledDisc}
\end{align}
where $t\in\mathbb{N}$ is the iteration number. The recursion is initialized with $x^{(0)}=x_{\text{\tiny max}}$. 
Since $f(g([0,x_{\text{\tiny max}}(\epsilon)])))\subset [0,x_{\text{\tiny max}}(\epsilon)]$, the initialization of \eqref{eqn:scalarDEuncoupledDisc} implies that $x^{(1)}\leq x^{(0)}=x_{\text{\tiny max}}$ and more generally $x^{(t+1)}\leq x^{(t)}$.
Thus $x^{(t)}$ will converge to a limiting value $x^{(\infty)}$ and this limit is a fixed point since $f$ and $g$ are continuous. 
The fixed points of the recursion \eqref{eqn:scalarDEuncoupledDisc} can be described as stationary points of 
a {\it single system} potential function $U_s$ defined as
\begin{align}
U_s(x)=xg(x;\epsilon)-G(x;\epsilon)-F(g(x;\epsilon);\epsilon),
\end{align}
where $F(x;\epsilon)=\int_{g(0;\epsilon)}^x\mathrm{d}s\,f(s;\epsilon)$ and $G(x;\epsilon)=\int_0^x\mathrm{d}s\,g(s;\epsilon)$. 
Without loss of generality, this function is normalized so that $U_s(x)=0$.

We define $x_{\text{\tiny good}}(\epsilon)$ as the fixed point of the recursion \eqref{eqn:scalarDEuncoupledDisc} that is reached with an 
initialization $x^{(0)} =0$. Furthermore, the {\it algorithmic threshold}\footnote{Here, we mean the algorithmic threshold of the message passing algorithm underlying 
the recursion.} is defined as 
\begin{align}
\epsilon_a =\sup\{\epsilon\mid x^{(\infty)} = x_{\text{\tiny good}}\}.
\end{align}
The monotonicity of $f$ and $g$ implies that, for $\epsilon <\epsilon_a$, the basin of attraction of $x_{\text{\tiny good}}$ is the whole interval $[0, x_{\max}(\epsilon)]$. Moreover $x_{\text{\tiny good}}$ is the unique stationary point of the potential function and is a minimum since it is an attractive fixed point. For $\epsilon > \epsilon_a$ we have $x^{(\infty)} \neq x_{\text{\tiny good}}$ and we set
$x^{(\infty)} = x_{\text{\tiny bad}}$ (where $x_{\text{\tiny bad}}$ depends on $\epsilon$). Note that this is an attractive fixed point and is thus a (local) minimum of $U_s(x)$. 
The two attractive fixed points are separated by at least one unstable fixed point $x_{\text{\tiny unst}}$ which is a local maximum of $U_s(x)$. We henceforth assume that there does not appear any other fixed point besides
$x_{\text{\tiny good}}$, $x_{\text{\tiny unst}}$, and $x_{\text{\tiny bad}}$. With this assumption in mind, we define the {\it energy gap}
as 
\begin{align}
\Delta E = U_s(x_{\text{\tiny bad}}) - U_s(x_{\text{\tiny good}}).
\end{align}
and the {\it potential threshold} as the unique value $\epsilon_{\text{\tiny pot}}$ such that $\Delta E =0$ (it can be shown that $\Delta E$ in non-increasing in $\epsilon$). 

The corresponding spatially coupled recursions are obtained by placing $\len+w$ replicas of the single system on the spatial positions $z\in\{-w+1,\dots,\len\}$ and coupling them with a uniform coupling window of size $w$. The coupled recursion takes the form
\begin{align}
x_z^{(t+1)}=\frac{1}{w}\sum\limits_{j=0}^{w-1} \, f\big(\frac{1}{w}\sum\limits_{k=0}^{w-1} \, g(x_{z-j+k}^{(t)};\epsilon);\epsilon \big).\label{eqn:DEcoupledDisc}
\end{align}
Motivated by the phenomenology observed in many examples (e.g. for the BEC or for compressve sensing), in order to study the stationary phase where a soliton appears, 
we fix the boundary conditions as $x_z^{(t)}=x_{\text{\tiny good}}$, for $z=\{-w+1,\dots,-1\}$ and all $t\in\mathbb{N}$. The initialization of the recursion is $x_z^{(0)}=x_{\text{\tiny bad}}$, for $z=\{0,\dots,\len\}$. 
The corresponding {\it potential functional} is given by
\begin{align}
U(\mathbf{x})=\sum_{z=-w+1}^{\len}&\big(
x_zg(x_z;\epsilon)-G(x_z;\epsilon)\big)
-\sum_{z=-w+1}^{\len}F\Big(\frac{1}{w}\sum_{i=0}^{w-1}g(x_{z+i};\epsilon);\epsilon\Big),
\end{align}
where $\mathbf{x}=(x_{-w+1},\dots,x_{\len})$. The fixed point equation \eqref{eqn:DEcoupledDisc} can be obtained by setting the derivative with respect to $\mathbf{x}$ of the potential $U_c(\mathbf{x})$ to zero. 

The spatially coupled recursions \eqref{eqn:DEcoupledDisc} display the {\it threshold saturation} property. Namely, for all $\epsilon<\epsilon_{\text{\tiny pot}}$ the fixed point $x_z^{(\infty)}$, $z=-w+1, \dots, L$, of the recursion \eqref{eqn:DEcoupledDisc} is equal to a constant profile $x_{\text{\tiny good}}$.
In the remainder of this section, we consider the range $\epsilon\in[\epsilon_{\text{\tiny BP}},\epsilon_{\text{\tiny MAP}}]$. It is for these values of the parameter $\epsilon$ that a soliton propagating at finite speed is observed, after a transient phase lasting only for a few iterations.
The soliton is a kink with a front at position $z_{\text{\tiny front}}$, making a quick transition of width $O(2w)$, between the two values $x_z^{(t)}\approx x_{\text{\tiny good}}$ for $z << z_{\text{\tiny front}}$ and $x_z^{(t)}\approx x_{\text{\tiny bad}}$ for $z >> z_{\text{\tiny front}}$. 

The simplest example to keep in mind for the setting described above, as well as for the next paragraph, 
is the case of LDPC$(\lambda, \rho)$ codes with transmission over the BEC$(\epsilon)$ where $f(x;\epsilon)=\epsilon\lambda(x)$ and $g(x;\epsilon) = \rho(x)$
and $U_s(x)$ is equal to \eqref{potential-BEC-first-time}. Here $\epsilon_a=\epsilon_{\text{\tiny BP}}$, $\epsilon_{\text{\tiny pot}} =\epsilon_{\text{\tiny MAP}}$, $x_{\text{\tiny good}}=0$ and $x_{\text{\tiny bad}}= x_{\text{\tiny BP}}$ is the non-trivial BP fixed point. 

\subsection{Continuum limit and velocity formula for scalar systems}\label{ssection:vScalarStatement}

We consider the system in the limit $\len\gg w \gg 1$ and formulate a continuum approximation.
The coupled recursion \eqref{eqn:DEcoupledDisc} becomes
\begin{align}\label{eqn:DEcoupledCont}
x(z,t+1)=\int_0^1\mathrm{d}u\, f\big(\int_0^1\mathrm{d}s\,g(x(z-u+s,t);\epsilon);\epsilon \big).
\end{align}
We take the boundary condition
$x(z,t)\to x_{\text{\tiny good}}$ when $z\to -\infty$ 
and $x(z,t)\to x_{\text{\tiny bad}}$ when $z\to+\infty$. This boundary condition captures the profiles obtained after the transient phase has passed, and is well adapted to the study of the soliton propagation.

\vskip 0.25cm 
\noindent{\bf Velocity formula for scalar systems.} As before, we assume that there exists a constant $v>0$ such that, for $t\to +\infty$ and $\vert z-vt\vert=O(1)$, 
the profile $x(z, t) \to X(z-vt)$, where $X(z)$ is independent of the initial condition $x(z, 0)$ and satisfies 
$\lim_{z\to -\infty}X(z) = x_{\text{\tiny good}}$, $\lim_{z\to +\infty}X(z) = x_{\text{\tiny bad}}$. 
Under this assumption, the velocity of the soliton is
\begin{align}\label{eqn:vScalarFormula}
v=\frac{\Delta E}{\int_{\mathbb{R}}\mathrm{d}z\,g'(X(z);\epsilon)X'(z)^2},
\end{align}
where the shape $X(z)$ satisfies
\begin{align}
X(z) - v X^\prime(z) = \int_0^1\mathrm{d}u\, f\big(\int_0^1\mathrm{d}s\,g(X(z-u+s);\epsilon);\epsilon\big)
\end{align}

\subsection{Derivation of the velocity formula \eqref{eqn:vScalarFormula}}\label{ssection:derivationscalar}

The derivation of \eqref{eqn:vScalarFormula} follows closely that in Section~\ref{ssection:vBMSderivation},
so we will be quite brief. The first step is to introduce 
 a continuum version of $U(\mathbf{x})$, which we call $\Delta\mathcal{U}(x(\cdot,\cdot))$. We define $x_0(z)$ as a static (time-independent) profile that satisfies the boundary conditions $x_0(z)\to x_{\text{\tiny good}}$ when $z\to -\infty$ 
and $x_0(z)\to x_{\text{\tiny bad}}$ when $z\to+\infty$ (for example, one may take a Heaviside step function). 
This is a reference profile in order to have well-defined integrals in the following expression
\begin{align*}
\Delta\mathcal{U}(x(\cdot,\cdot))  =  \int_\mathbb{R}\mathrm{d}z\,\Big[&\{x(z,t)g(x(z,t);\epsilon) - G(x(z,t);\epsilon)
-F\Big(\int_0^1\mathrm{d}u\,g(x(z-u,t);\epsilon);\epsilon\big)\}
\nonumber \\ &
- \{x_0(z)g(x_0(z);\epsilon)
- G(x_0(z);\epsilon)
- F\big(\int_0^1\mathrm{d}u\,g(x_0(z-u);\epsilon);\epsilon \big)\}\Big].
\end{align*}
As long as $x(z,t)$ and $x_0(z)$ converge to their limiting values fast enough, the integrals over the spatial axis are well defined. 
Evaluating the functional 
derivative\footnote{Defined as $\lim_{\gamma\to 0}\gamma^{-1}(\Delta\mathcal{U}(x(\cdot ,\cdot) +\gamma\eta(\cdot,\cdot)) 
-\Delta\mathcal{U}(x(\cdot ,\cdot)))$}  
of $\Delta\mathcal{U}[x(\cdot,\cdot);\epsilon]$ in an 
arbitrary direction $\eta(\cdot,\cdot)$, we find that \eqref{eqn:DEcoupledCont} is equivalent to a gradient descent equation
\begin{align}
\int_\mathbb{R}\mathrm{d}z\, & g'(x(z,t);\epsilon)\big(x(z,t+1)-x(z,t)\big)\eta(z,t)
= - \frac{\delta\Delta\mathcal{U}}{\delta X}[\eta(z,t)]
\label{gradient-scalar}
\end{align}
Now we use the ansatz $x(z,t)\to X(z-vt)$ and apply \eqref{gradient-scalar} for the 
special direction $\eta(z,t)= X^\prime(z-vt)$. Using also the approximation 
$X(z-vt) \approx X(z) -v X^\prime(z)$ for small $v$ we get (after a change of variables 
$z\to z+vt$)
\begin{align}
v\int_\mathbb{R}\mathrm{d}z\,X'(z)^2g'(X(z);\epsilon) = \frac{\delta\Delta\mathcal{U}}{\delta x}[X^\prime(z)].
\end{align}
We
then proceed to compute the right-hand side of \eqref{gradient-scalar}. We split the potential functional into two parts: the ``single system potential'' $\mathcal{U}_s(x(\cdot,\cdot))$ 
that remains if we ignore the coupling effect,
and the ``interaction potential'' $\mathcal{U}_i(x(\cdot,\cdot))$ that captures the effect of coupling. That is, 
$\Delta\mathcal{U} = \mathcal{U}_s +\mathcal{U}_i$, with 
\begin{align*}
\mathcal{U}_s(x(\cdot,\cdot)) =  \int_\mathbb{R}\mathrm{d}z\,\Big[&\{x(z,t)g(x(z,t);\epsilon)
-G(x(z,t);\epsilon)-F(g(x(z,t);\epsilon)\}
\nonumber \\ &
-\{x_0(z)g(x_0(z);\epsilon) 
- G(x_0(z);\epsilon)-F(g(x_0(z);\epsilon);\epsilon)\}\Big],\\
\mathcal{U}_i(x(\cdot,\cdot)) =  \int_\mathbb{R}\mathrm{d}z\,\Big[&\{F(g(x(z,t);\epsilon);\epsilon)
-F(\int_0^1\mathrm{d}u\,g(x(z-u,t);\epsilon)\}
\nonumber \\ &
-\{F(g(x_0(z);\epsilon);\epsilon) 
- F\Big(\int_0^1\mathrm{d}u\,g(x_0(z-u);\epsilon);\epsilon \big)\}\Big].
\end{align*}
The computation of each functional derivative at $x(z,t)\to X(z-vt)$ in the direction $X^\prime(z-vt)$ yields  
\begin{align}
\frac{\delta\mathcal{U}_s}{\delta X}  [X'(z)] 
&=\int_\mathbb{R}\mathrm{d}z\,X'(z)\Big(X(z)g'(X(z);\epsilon)
-g'(X(z);\epsilon)f(g(X(z);\epsilon);\epsilon)\Big) 
\nonumber \\
&=\int_\mathbb{R}\mathrm{d}z\,\frac{\mathrm{d}}{\mathrm{d}z}\Big\{X(z)g(X(z);\epsilon) 
-G(X(z);\epsilon)-F(g(X(z);\epsilon);\epsilon)\Big\}
\nonumber \\
&=\Bigl[U_s(X(z))\Bigr]_{-\infty}^{+\infty}
\nonumber \\ &
=U_s(x_{\text{\tiny bad}};\epsilon)-U_s(x_{\text{\tiny good}};\epsilon).
\label{firstderivativescalar}
\end{align}
and 
\begin{align}
\frac{\delta\mathcal{U}_i}{\delta X}  [X'(z)] 
& = \int_{\mathbb{R}}\mathrm{d}z X^\prime(z)\Big\{f(g(X(z);\epsilon);\epsilon) g^\prime(X(z)) 
- f(\int_0^1 \mathrm{d}u g(X(z-u);\epsilon);\epsilon)\int_0^1 \mathrm{d}u g^\prime(X(z-u);\epsilon)\Big\}
\nonumber \\
&=\int_\mathbb{R}\mathrm{d}z\frac{\mathrm{d}}{\mathrm{d}z}\Big\{F(g(X(z);\epsilon);\epsilon) 
-F\big(\int_0^1\mathrm{d}u\,g(X(z-u);\epsilon);\epsilon\big)\Big\}
\nonumber \\
&=0.
\label{secondderivative-scalar}
\end{align}
Replacing \eqref{firstderivativescalar} and \eqref{secondderivative-scalar} in \eqref{gradient-scalar}, we obtain the 
velocity formula \eqref{eqn:vScalarFormula}.

\subsection{Generalized LDPC (GLDPC) Codes}\label{ssection:GLDPC}

A GLDPC code is a code represented by a bipartite graph, such that the rules of 
the check nodes do not depend on parity (as do usual LDPC codes) but on a primitive BCH code. An attractive property of BCH codes is that they can be designed to correct a chosen number of errors. For instance, one can design a BCH code so that it corrects all patterns of at most $e$ erasures on the BEC, and all error patterns of weight at most $e$ on the binary symmetric channel (BSC). We consider a GLDPC code with degree-2 variable nodes and degree-$n$ check nodes whose rules are given by a primitive BCH code of blocklength $n$. 

We give a short description of a BCH code of blocklength $n$ and minimum distance $d=2e+1$
(see \cite{pless2011introduction} for more details. A BCH code is a cyclic code over a finite field GF($b^\beta$) where $b$ is a prime power and $\beta$ is an integer. Let $a$ be a primitive element of GF($b^\beta$). Then each element of GF($b^\beta$) can be written in the form $a^i$, $i\in\mathbb{N}$. For each element $a^i$ we can define a minimal polynomial $m_i(\cdot)$ which is the monic polynomial over GF($b$) with smallest degree. The generator polynomial $\theta(\cdot)$ over GF($b$) of the BCH code is defined as the least common multiple of $m_1(\cdot),\dots,m_d(\cdot)$.

Consider transmission on the BEC or BSC and denote by $\epsilon$ the channel parameter. The density evolution recursions have been derived in \cite{jian2012approaching} for both channels, based on a bounded distance decoder for the BCH code. For $n$ and $e$ fixed, we can write the update equations of the message passing algorithm as \eqref{eqn:scalarDEuncoupledDisc} with
\begin{align*}
\begin{cases}
f(x;\epsilon)=\epsilon x,\\
g(x)=\sum_{i=e}^{n-1}{{n-1}\choose{i}}x^i(1-x)^{n-i-1}.
\end{cases}
\end{align*}
Here, we have $\epsilon_{\text{\tiny max}}=x_{\text{\tiny max}}=y_{\text{\tiny max}}=1$. Moreover, one checks easily by differentiation (with respect to $x$) that 
the potential function $U_{\text{\tiny GLDPC}}(x)$ of the system is given by
\begin{align*}
U_{\text{\tiny GLDPC}}(x)=\frac{e}{n}g(x;\epsilon)-\frac{g'(x;\epsilon)}{n\,x(1-x)}-\frac{\epsilon}{2}g^2(x;\epsilon).
\end{align*}
For numerical implementation purposes, it is useful to note that
\begin{align*}
g'(x)=\frac{x^{e-1}(1-x)^{n-e-1}}{B(e,n-e)},
\end{align*}
where $B(a,b)=\frac{(a-1)!(b-1)!}{(a+b-1)!}$ denotes the Beta function, and that $g(x)$ is equal to the regularized incomplete Euler Beta function so that
\begin{align*}
g(x)=\frac{1}{B(e,n-e)}\int_0^x\mathrm{d}s\,s^{e-1}(1-s)^{n-e-1}.
\end{align*}
This potential has $x=0$ as a trivial stationary point (equivalently, this is a  trivial fixed point of DE as can be seen from the expressions of $f$ and $g$) and develops a non-trivial minimum at $x_{\text{\tiny BP}}\neq 0$ for $\epsilon >\epsilon_{\text{\tiny BP}}$. Note that $\epsilon_{\text{\tiny BP}}$ is found as usual as the first horizontal inflexion point. The 
MAP threshold is given by $\epsilon_{\text{\tiny MAP}}$ such 
that $U_{\text{\tiny GLDPC}}(x_{\text{\tiny BP}}) = 
U_{\text{\tiny GLDPC}}(0) =0$.

\begin{figure}
\centering
\includegraphics[draft=false,scale=0.27]{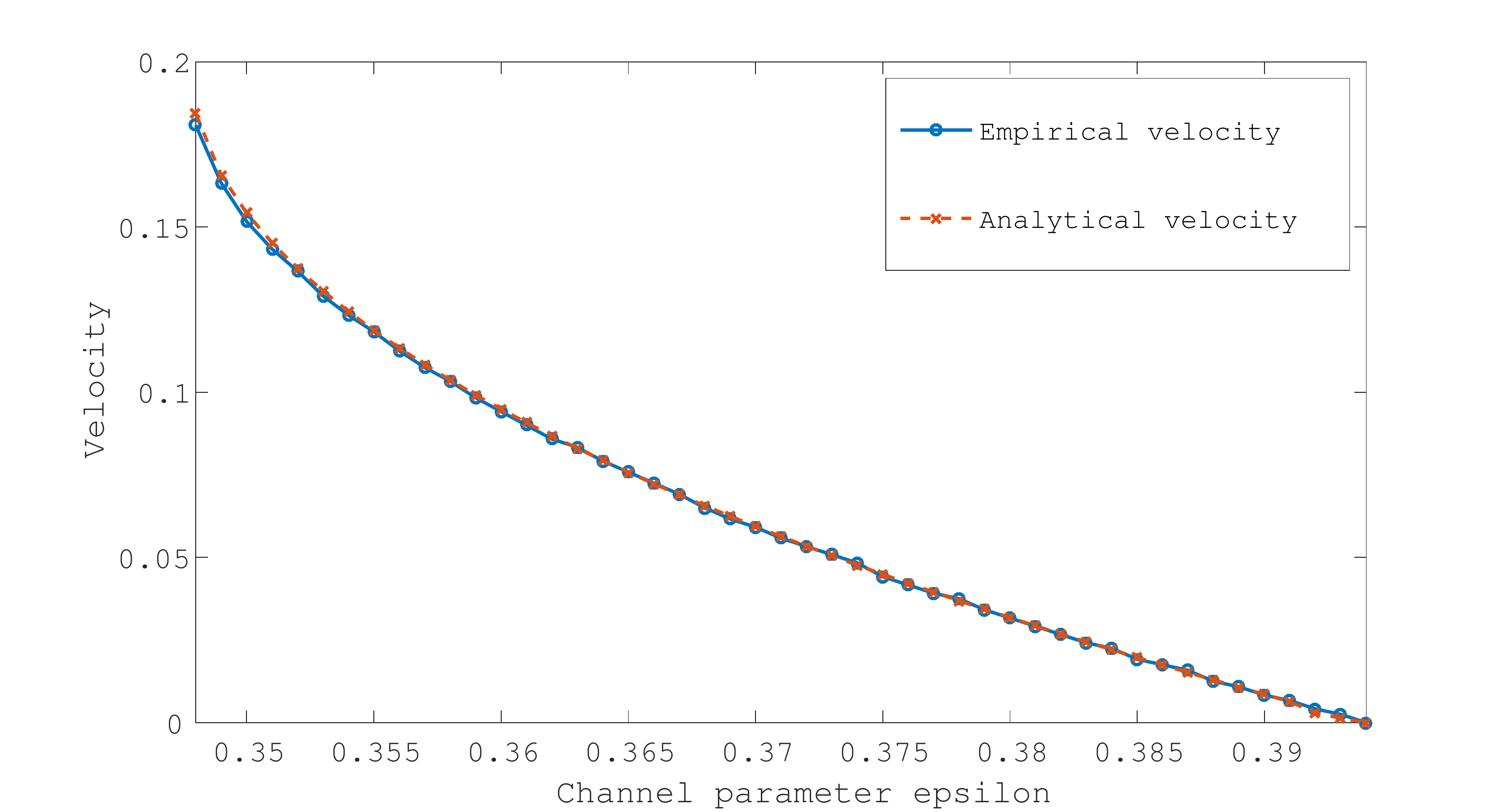}
\caption{We consider a GLDPC code with $n=15$ and $e=3$, with spatial length $L+w=500$ and uniform coupling window with $w=3$. We plot the normalized velocities $v_{\text{\tiny GLDPC}}$ and $v_e$ as a function of the channel parameter $\epsilon$ when $\epsilon$ is between the BP threshold $\epsilon_s=\epsilon_{\text{\tiny BP}}= 0.348$ and the potential threshold $\epsilon_c=\epsilon_{\text{\tiny MAP}}\approx 0.394$.}
\label{fig:velocitiesGLDPC}
\end{figure}

The formula for the velocity of the soliton appearing for coupled GLDPC codes is found from \eqref{eqn:vScalarFormula}. The energy gap for $\epsilon_{\text{\tiny BP}} \leq \epsilon \leq \epsilon_{\text{\tiny MAP}}$ 
is now $\Delta E=U_{\text{\tiny GLDPC}}(x_{\text{\tiny BP}})-U_{\text{\tiny GLDPC}}(0)$.
Figure~\ref{fig:velocitiesGLDPC} shows the velocities (normalized by $w$) for the spatially coupled GLPDC code with $n=15$ and $e=3$, when the coupling parameters satisfy $\len+w=500$ and $w=3$. We plot the velocities for $\epsilon\in[\epsilon_{\text{\tiny BP}},\epsilon_{\text{\tiny MAP}}]=[0.348,0.394]$. We observe that the formula for the velocity provides a very good estimation of the empirical velocity $v_e$.

\subsection{Compressive Sensing}\label{ssection:CS}

Let $\mathbf{s}$ be a length-$n$ signal vector where the components are i.i.d. copies of a random variable $S$. We take $m$ linear measurements of the signal and assume that the measurement matrix has i.i.d Gaussian elements distributed like $\mathcal{N}(0, 1/\sqrt n)$. We define $\delta =m /n$ as the measurement ratio and fix it to a constant value when $n\to\infty$. The relation between $\delta$ and the parameter $\epsilon$ defined in Section~\ref{ssection:vScalarNotation} is $\epsilon = \delta^{-1}$. 
We assume that the power of the variable $S$ is normalized to 1; that is, $\mathbb{E}[S^2]=1$. We also assume that each component of the signal $\mathbf{s}$ is corrupted by independent Gaussian noise of variance $\sigma^2=1/\mathtt{snr}$.
To recover $\mathbf{s}$ one implements the so-called approximate message passing (AMP) algorithm. 

It is well-known that the analysis of the AMP algorithm is given by state evolution \cite{donoho2012information}. 
Let $Y =\sqrt{\mathtt{snr}}\,S+Z$ where $Z\sim \mathcal{N}(0,1)$, let
 $\hat S(Y)=\mathbb{E}_{S|Y}[S|Y]$ the minimum mean square estimator, and set
 \begin{align*}
\mathtt{mmse}(\mathtt{snr})=\mathbb{E}_{S,Y}[(S-\hat S(Y,\mathtt{snr}))^2],
\end{align*} 
for the $\mathtt{mmse}$ function.
The state evolution equations (which track the mean squared error of the AMP estimate) then correspond to the recursion \eqref{eqn:scalarDEuncoupledDisc} with
\begin{align*}
\begin{cases}
f(x,\delta)=\mathtt{mmse}(\mathtt{snr}-x),\\
g(x,\delta)=\mathtt{snr}-\frac{1}{\frac{1}{\mathtt{snr}}+\frac{x}{\delta}}.
\end{cases}
\end{align*}
Here $x$ is interpreted as the mean square error predicted by the AMP estimate of the signal. State evolution is 
initialized with $x=1$ which corresponds to no knowledge about the signal. 
We will take $\delta$ as the control parameter.  Note that we
 have $\delta_{\text{\tiny max}}=1$, $x_{\text{\tiny max}}=\mathtt{mmse}(0)$, $y_{\text{\tiny max}}=g(x_{\text{\tiny max}})$. The potential function is equal to 
\begin{align*}
U_{\text{\tiny CS}}&(x)
=-\frac{x}{\frac{1}{\mathtt{snr}}+\frac{x}{\delta}}+\delta\ln\Big(1+\frac{x \, \mathtt{snr}}{\delta}\Big)
-2I\Big(S;\sqrt{\mathtt{snr}}S+Z\Big)+2I\Big(S;\sqrt{\frac{1}{\frac{1}{\mathtt{snr}}+\frac{x}{\delta}}}S+Z\Big),
\end{align*}
where $I(A;B)$ is the mutual information between two random variables $A$ and $B$.
To check that this potential gives back the correct state evolution equation as a stationarity condition, we simply differentiate it
with respect to $x$  thanks to the well known 
relation $\frac{\mathrm{d}}{\mathrm{d}\mathtt{snr}}I(S;\sqrt{\mathtt{snr}}S+Z) = \frac{1}{2}\mathtt{mmse}(\mathtt{snr})$. 

\begin{figure}
\centering
\includegraphics[draft=false,scale=0.27]{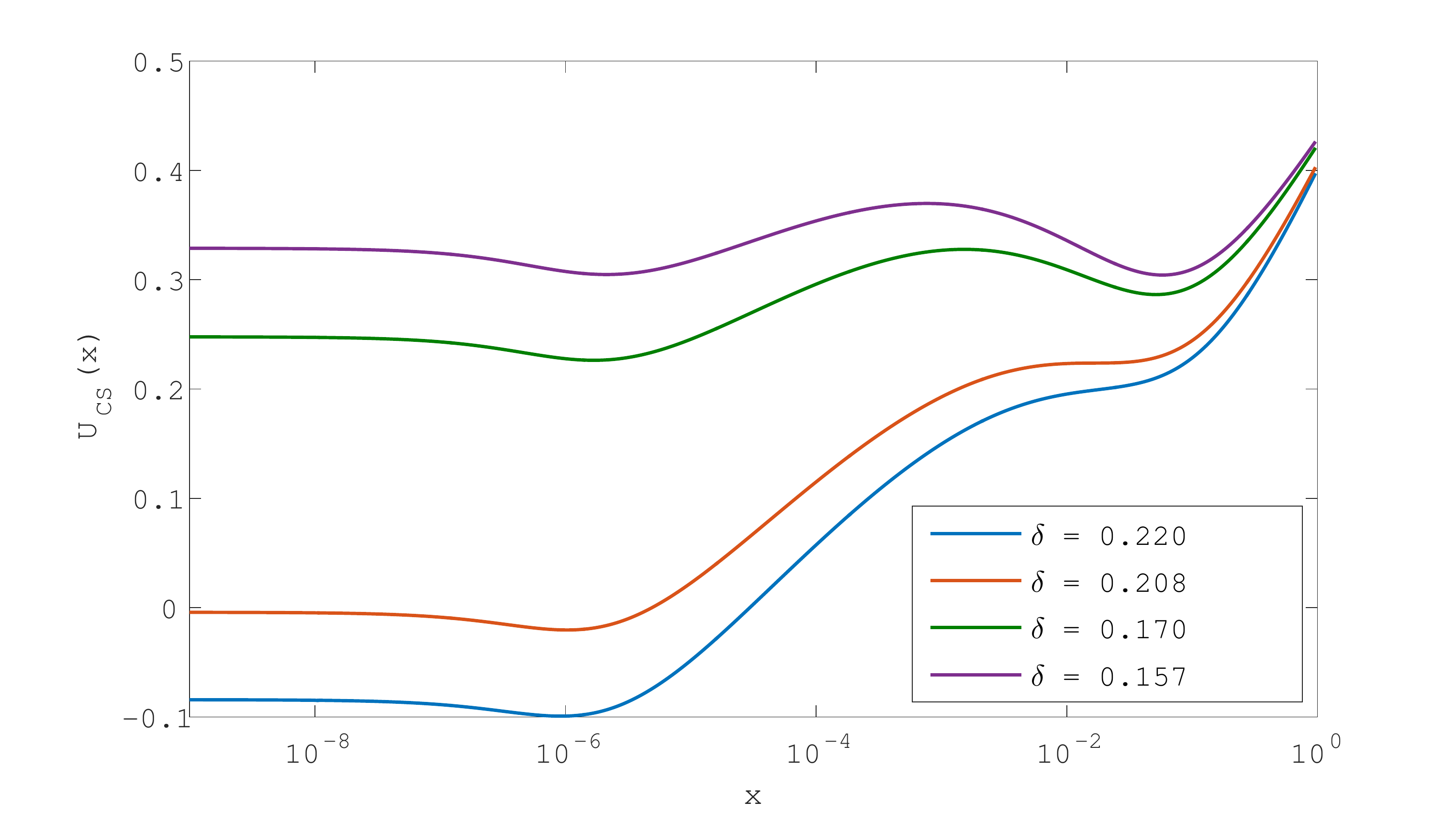}
\caption{Potential function for compressive sensing with a Gaussian-Bernoulli prior. The sparsity parameter is $\rho =0.1$ and the signal to noise ratio $\mathtt{snr}=10^5$. We show the potential for several values of  the measurement fraction $\delta$. For $\delta > \delta_{\text{\tiny AMP}}$ we have a single minimum $x_{\text{\tiny good}}$. At $\delta_{\text{\tiny AMP}}= 0.208$ there is a horizontal inflexion point and for smaller 
measurement fractions a second minimum $x_{\text{\tiny bad}}$ appears. At $\delta_{\text{\tiny opt}}=0.157$ the gap $\Delta E$ vanishes.}
\label{fig:potential-comp-sensing}
\end{figure}

To illustrate the potential function in a concrete case we take 
the Bernoulli-Gaussian distribution as the prior distribution over the signal components
\begin{align*}
q_0(s)=(1-\rho)\delta(s)+\rho\frac{e^{-s^2/2}}{\sqrt{2\pi}},
\end{align*}
Figure~\ref{fig:potential-comp-sensing} shows the potential function for $\rho=0.1$, $\mathtt{snr}=10^5$, and several values of $\delta$ (the measurement fraction). We observe that, for $\delta > \delta_{\text{\tiny AMP}} = 0.208$, there is 
a unique minimum $x_{\text{\tiny good}}$ which is a fixed point of state evolution when it is initialized to $x=1$. In this phase, there are enough measurements so that the reconstruction of the signal is good and the mean square error is small. At 
$\delta_{\text{\tiny AMP}} = 0.208$ a horizontal inflexion point develops in the potential function. For $\delta<\delta_{\text{\tiny AMP}}$ a second minimum appears at a higher mean square error $x_{\text{\tiny bad}}$ and the reconstruction of the AMP algorithm is bad. The optimal threshold corresponding to the minimum mean square error estimator is found 
when $\delta$ is such that the two minima of the potential are at the same height, namely $\delta_{\text{\tiny opt}}$ is given 
by the solution of the equation
$U_{\text{\tiny SC}}(x_{\text{\tiny bad}}) = U_{\text{\tiny SC}}(x_{\text{\tiny good}})$. For our example one finds $\delta_{\text{\tiny opt}} = 0.157$. This threshold is reached by the AMP algorithm on the spatially coupled system. 

\begin{figure}
\centering
\includegraphics[draft=false,scale=0.27]{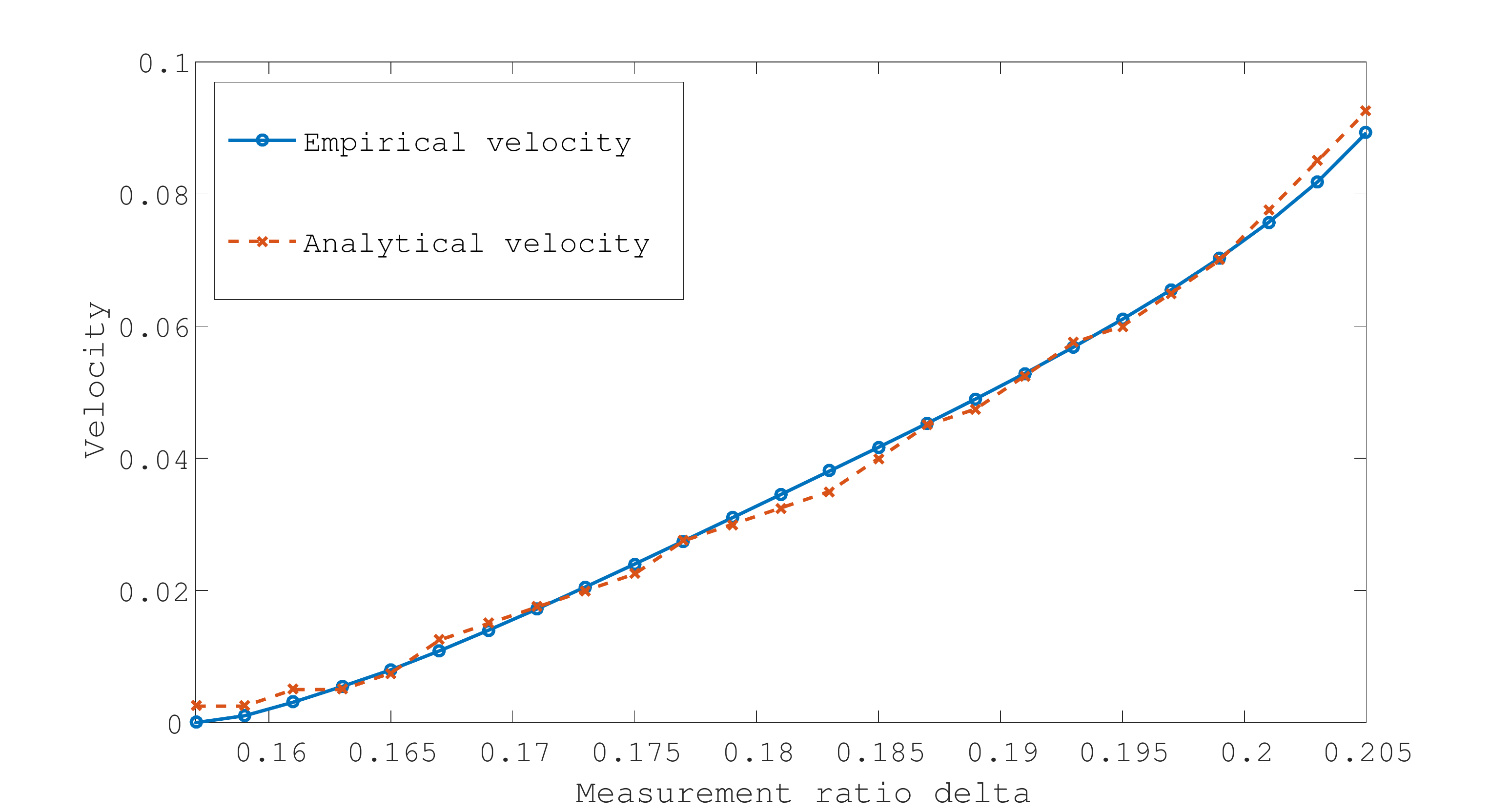}
\caption{We consider the compressive sensing problem with $\mathtt{snr}=10^5$ and Gaussian-Bernoulli prior for the signal components with sparsity parameter $\rho=0.1$. We have $\len+w=250$ and uniform coupling window with $w=4$. We plot the normalized velocities $v_{\text{\tiny CS}}$ and $v_e$ as a function of the measurement fraction $\delta$ when $\delta$ is between the potential threshold $\delta_{opt}=0.157$ and $\delta_{\text{\tiny AMP}}= 0.208$.}
\label{fig:velocitiesCS}
\end{figure}

Fix $\delta\in[\delta_{\text{\tiny opt}},\delta_{\text{\tiny AMP}}]=[0.157,0.208]$. In this regime 
spatially coupled state evolution develops a soliton which represents the profile of mean square error
along the spatial direction. 
The formula for the velocity $v_{\text{\tiny CS}}$ of this soliton is obtained from \eqref{eqn:vScalarFormula} 
where the energy gap is now $\Delta E=U_{\text{\tiny SC}}(x_{\text{\tiny bad}})-U_{\text{\tiny SC}}(x_{\text{\tiny good}})$. 
Figure~\ref{fig:velocitiesCS} shows the velocities (normalized by $w$) for the spatially coupled compressive sensing system with $\mathtt{snr}=10^5$, $\rho=0.1$ and with the coupling parameters satisfy $\len+w=250$ and $w=4$. %
We plot in Figure~\ref{fig:velocitiesCS} the velocities for $\delta\in[\delta_{\text{\tiny opt}},\delta_{\text{\tiny AMP}}]=[0.157,0.208]$. It is clear that the formula for the velocity provides a good estimation of the empirical velocity $v_e$. 

\section{Conclusion}\label{section:conclusion}

Our formulas for the velocities of the solitons that appear in spatially coupled codes for BMS channels and for general 
spatially coupled scalar systems (e.g. compressive sensing) rest on the approximation of the discrete system 
by a continuum one. We believe that this approximation becomes exact in an asymptotic limit of infinite spatial length and window size
(keeping the order $\len \gg w\gg 1$). It is an interesting open problem to quantify the quality of this approximation already for $\len$ infinite and 
$w$ finite but large. The numerical results tend to indicate that the approximation is 
already quite good for small values of $w$, when it is of the order of a few positions. 

Another important and interesting open question concerns the proof of the ansatz, namely proving that the shape of the soliton is unique 
and independent of the initial condition on the profile. Settling this question would show that the velocity of the soliton is independent 
of the size of the seed that initiates decoding or signal reconstruction at the boundaries.  

The formulas for the velocity involve the whole shape of the soliton in the denominator. For optimization purposes it 
would be desirable to have a good approximation (or bound) on the denominator that would involve only primitive quantities 
related to the underlying uncoupled ensemble (such as the degree distributions, the single system potential, etc). Such an approximation scheme 
has been proposed in \cite{rafah-nicolas-ISIT2016} for the special case of the transmission over the BEC where it works quite well
 close to the MAP threshold. It would be desirable to find an extension to the more general situations considered in the present paper. 

\appendices

\section{Functional derivatives}\label{appendix}

In this appendix we derive equations \eqref{funct-derivative-conti} and \eqref{driva-Ws}. 

\subsection{Derivation of Equ. \eqref{funct-derivative-conti}}

We calculate the functional derivative of $\Delta\mathcal{W}(\mathtt{x})$
in a direction $\mathtt{\eta}(z, t)$ as follows.
\begin{align*}
 &\frac{\delta \Delta \mathcal{W}}{\delta \mathtt{x}}[\mathtt{\eta}(z,t)]
 =
\frac{\partial}{\partial \gamma}\Delta \mathcal{W}(\mathtt{x}+\gamma \mathtt{\eta})\Big|_{\gamma=0} 
 = \frac{\partial}{\partial \gamma}\int_\mathbb{R} \mathrm{d}z \, P(z,\mathtt{x}+\gamma \mathtt{\eta})\Big|_{\gamma=0},
\end{align*}
where the function $P(\cdot,\cdot)$ is defined in \eqref{eqn:coupledPotCont}. Then, taking the derivative with respect to $\gamma$ yields
\begin{align*}
 \int_\mathbb{R} \mathrm{d}z \, \Big\{&H(\rho^\boxast(\mathtt{x}(z,t))\boxast\mathtt{\eta}(z,t))
+ H(\rho^{\prime\boxast}(\mathtt{x}(z,t))\boxast\mathtt{\eta}(z,t)) 
- H(\rho^\boxast(\mathtt{x}(z,t))\boxast\mathtt{\eta}(z,t))\\
&- H(\mathtt{x}(z,t)\boxast \rho^{\prime\boxast}(\mathtt{x}(z,t))\boxast\mathtt{\eta}(z,t))
-H\Big(\mathtt{c}(z)\varoast\lambda^\varoast\Big(\int_0^1\mathrm{d}s \, \rho^\boxast(\mathtt{x}(z+s,t)) \Big)\\
& \qquad \qquad \varoast\Big[\int_0^1\mathrm{d}u \, \rho^{\prime\boxast}(\mathtt{x}(z+u,t))\boxast\mathtt{\eta}(z+u,t)\Big] \Big)\Big\}.
\end{align*}
We notice that the first and third terms in the integral cancel out due to the commutativity of the operator $\boxast$. By rearranging the averaging functions in the last term, we obtain
\begin{align*}
 \int_\mathbb{R} & \mathrm{d}z \, \Big\{H(\rho^{\prime\boxast}(\mathtt{x}(z,t))\boxast\mathtt{\eta}(z,t))
- H(\mathtt{x}(z,t)\boxast \rho^{\prime\boxast}(\mathtt{x}(z,t))\boxast\mathtt{\eta}(z,t))\\
&-H\Big(\int_0^1\mathrm{d}u \,\mathtt{c}(z-u)\varoast\lambda^\varoast\Big(\int_0^1\mathrm{d}s \, \rho^\boxast(\mathtt{x}(z-u+s,t)) \Big)
 \varoast\Big[ \rho^{\prime\boxast}(\mathtt{x}(z,t))\boxast\mathtt{\eta}(z,t)\Big] \Big)\Big\}.
\end{align*}
By noticing that $\mathtt{y} = \int_0^1\mathrm{d}u \,\mathtt{c}(z-u)\varoast\lambda^\varoast(\int_0^1\mathrm{d}s \, \rho^\boxast(\mathtt{x}(z-u+s,t)))$ is a probability measure and $\mathtt{a} = \rho^{\prime\boxast}(\mathtt{x}(z,t))\boxast\mathtt{\eta}(z,t)$ is a difference of probability measures, we can use the second duality rule in \eqref{eqn:operatorProp2} $H(\mathtt{y}\boxast\mathtt{a})+H(\mathtt{y}\varoast\mathtt{a})=H(\mathtt{a})$ to rewrite the above as, freely using the commutativity of $\boxast$,
\begin{align*}
 \int_\mathbb{R}  \mathrm{d}z \, \Big\{- H(\mathtt{x}(z,t)\boxast \rho^{\prime\boxast}(\mathtt{x}(z,t))\boxast\mathtt{\eta}(z,t))
 +H\Big(\Big[\int_0^1\mathrm{d}u \,\mathtt{c}(z-u)&\varoast\lambda^\varoast\Big(\int_0^1\mathrm{d}s \, \rho^\boxast(\mathtt{x}(z-u+s,t)) \Big)\Big]\\
&  \boxast \rho^{\prime\boxast}(\mathtt{x}(z,t))\boxast\mathtt{\eta}(z,t)\Big)\Big\}\\
 = \int_\mathbb{R}  \mathrm{d}z \, \rho^{\prime\boxast}(\mathtt{x}(z,t))\boxast\mathtt{\eta}(z,t)
\boxast\Big(\int_0^1\mathrm{d}u \,\mathtt{c}(z-u)\varoast\lambda^\varoast\Big(\int_0^1\mathrm{d}s \, \rho^\boxast&(\mathtt{x}(z-u+s,t)) \Big)
- \mathtt{x}(z,t) \Big).
\end{align*}

\subsection{Derivation of Equ. \eqref{driva-Ws}}

We calculate the functional derivative of $\mathcal{W}_s(\mathtt{X})$
in the direction $\mathtt{X}'(z)$ as follows.
\begin{align*}
&\frac{\delta \mathcal{W}_s}{\delta \mathtt{X}}[\mathtt{X}'(z)] 
= \frac{\partial}{\partial \gamma}\mathcal{W}_s(\mathtt{X}+\gamma\mathtt{X}')\Big|_{\gamma=0}
= \frac{\partial}{\partial \gamma} \int_\mathbb{R} \mathrm{d}z \, P_s(z,\mathtt{X}+\gamma\mathtt{X}')\Big|_{\gamma=0} \\
& = \int_\mathbb{R} \mathrm{d}z \, \Big\{H(\rho^\boxast (\mathtt{X}(z))\boxast\mathtt{X}'(z))+H(\rho^{\prime\boxast} (\mathtt{X}(z))\boxast\mathtt{X}'(z))
- H(\rho^\boxast (\mathtt{X}(z))\boxast\mathtt{X}'(z))\\
&\qquad\qquad- H(\mathtt{X}(z) \boxast \rho^{\prime\boxast} (\mathtt{X}(z))\boxast\mathtt{X}'(z))
- H(\mathtt{c}(z) \varoast \lambda^\varoast(\rho^\boxast(\mathtt{X}(z))) 
\varoast [\rho^{\prime\boxast}(\mathtt{X}(z))\boxast\mathtt{X}'(z)]) \Big\}.
\end{align*}
We notice here that on the right side of the last equality, the first and third terms under the integral cancel out. Using the second duality rule in \eqref{eqn:operatorProp2}, and noticing that $\mathtt{X}(z)$ is a probability measure and $\rho^{\prime\boxast} (\mathtt{X}(z))\boxast\mathtt{X}'(z)$ is a difference of probability measures, we can rewrite the functional derivative as
\begin{align*}
&\int_\mathbb{R} \mathrm{d}z \, \Big\{H(\mathtt{X}(z) \varoast [\rho^{\prime\boxast} (\mathtt{X}(z))\boxast\mathtt{X}'(z)]) 
- H(\mathtt{c}(z) \varoast \lambda^\varoast(\rho^\boxast(\mathtt{X}(z))) 
\varoast [\rho^{\prime\boxast}(\mathtt{X}(z))\boxast\mathtt{X}'(z)]) \Big\}.
\end{align*}

%

\section{Derivation of expression \eqref{final-for-ga-velocity}}\label{appendixTaylor}

Our goal in this appendix is to show that \eqref{entropies-GA-delta2} 
reduces to \eqref{final-for-ga-velocity} when 
$\delta\to 0$.  We first reorganize \eqref{entropies-GA-delta2}  as follows (up to multiplication by $1/\delta^2$)
\begin{align*}
\psi\Big( & (r-2)\psi^{-1}(1-p(z))+2\psi^{-1}(1-p(z+\delta))\Big)
-2\psi\Big((r-2)\psi^{-1}(1-p(z))+\psi^{-1}(1-p(z+\delta))
\nonumber \\ & +\psi^{-1}(1-p(z))\Big)
+\psi\Big((r-2)\psi^{-1}(1-p(z))+2\psi^{-1}(1-p(z))\Big).
\end{align*}
When we Taylor expand each entropy $\psi (\cdots)$ 
around 
$$
(r-2)\psi^{-1}(1-p(z))
$$ 
to second order we observe that the first order terms cancel and what remains is
\begin{align*}
-\frac{1}{2}\psi''\big( & (r-2)\psi^{-1}(1-p(z))\big)\Big\{4\psi^{-1}(1-p(z+\delta))^2
-4\psi^{-1}(1-p(z))^2\\
&-2\bigl(\psi^{-1}(1-p(z+\delta))-\psi^{-1}(1-p(z)) \big)^2\Big\}.
\end{align*}
This is equal to 
\begin{align*}
- \psi''\big( & (r-2)\psi^{-1}(1-p(z))\big)
\times \Big(\psi^{-1}(1-p(z+\delta))-\psi^{-1}(1-p(z))\Big)^2.
\end{align*}
Next, we write $\psi^{-1}(1-p(z+\delta))$ as 
$$
\psi^{-1}(1-p(z)-(p(z+\delta)-p(z)))
$$ 
and Taylor expand $\psi^{-1}(\cdots)$ around $1-p(z)$ to obtain
\begin{align*}
- &(p(z+\delta)-p(z))^2 ((\psi^{-1})^{\prime}(1-p(z)))^2
\times
\psi''\big((r-2)\psi^{-1}(1-p(z))\big).
\end{align*}
Multiplying by $1/\delta^2$, taking the limit 
$\delta\to 0$, and using the relation $(\psi^{-1})'(\cdots)=1/(\psi'(\psi^{-1}(\cdots)))$ we obtain
\begin{align*}
-\Bigg(\frac{\mathrm{d}p(z)}{\mathrm{d}z}\Bigg)^2\frac{\psi''\big((r-2)\psi^{-1}(1-p(z))\big)}{\Big(\psi'(\psi^{-1}(1-p(z)))\Big)^2}.
\end{align*}

\section*{Acknowledgment}

The authors would like to thank Ruediger Urbanke and Markus Stinner for discussions and suggestions on applications to finite-length scaling laws, and Tongxin Li for interactions on compressive sensing during his summer internship in EPFL.

\ifCLASSOPTIONcaptionsoff
  \newpage
\fi

\bibliographystyle{IEEEtran}
\bibliography{BIBfile}

\end{document}